\documentclass[twocolumn,final,useamsfonts]{pasj02}

\usepackage{booktabs}
\usepackage{array}
\usepackage{needspace}
\usepackage{url}
\usepackage{xcolor}
\usepackage{placeins}
\newif\ifrevisionmarked
\revisionmarkedfalse
\ifdefined\RevisionClean\revisionmarkedfalse\fi
\newcommand{\revised}[1]{{#1}}
\newcommand{\rtwo}[1]{{\ifrevisionmarked\color{red}\fi#1}}
\providecommand{\href}[2]{#2}
\providecommand{\nolinkurl}[1]{\texttt{#1}}
\newcommand{\PasjBibDoi}[1]{doi:\,\href{https://doi.org/#1}{\nolinkurl{#1}}}
\newcommand{\PasjBibEprint}[1]{\nolinkurl{#1}}
\newcommand{\PasjBibArxiv}[1]{arXiv:\,\href{https://arxiv.org/abs/#1}{\nolinkurl{#1}}}

\jyear{2026}
\Received{}
\Accepted{}

\begin{document}

\title{A source-plane reassessment of the $z=6$--8 ultraviolet luminosity
function behind Abell 2744: no requirement for additional faint-end curvature}

\author{Xuheng \textsc{Ma},\altaffilmark{1}\altemailmark
\email{u3014707@connect.hku.hk}}
\altaffiltext{1}{Department of Physics, The University of Hong Kong,
Pokfulam Road, Hong Kong SAR, China}


\maketitle

\begin{abstract}
\revised{We reassess the rest-frame ultraviolet luminosity function at
$6\leq z<8$ behind Abell~2744 using public UNCOVER data.}
\revised{Seven-band artificial sources calibrate selection on the observed
galaxies' fixed-aperture scale. The source-plane response maps total luminosity
into measured bins, accounting for counterimages, flux errors, redshift
migration and the magnification cut. Of 184 registry-de-duplicated sources, 114 in
B6 ($6\leq z<7$) and 37 in B7 ($7\leq z<8$) enter the common fitted range of four
compact-profile/SED calibrations. Their five-per-cent relative-peak response
boundaries reach
approximately $M_{1500}=-16.4$ to $-16.7$ and $-17.0$ to $-17.1$,
respectively. Joint Poisson fits give curved-minus-Schechter
$\Delta\mathrm{AIC}=+4.00$ and $+3.87$ to $+3.95$. The B6 value is the
penalty for two extra parameters with no likelihood gain. A matched-area
GLAFIC comparison leaves this ranking unchanged.
We tested curvature in B6 or B7 that begins at $M_{\rm t}=-18.5$ or $-17.5$ and
lowers the density by 21 or 60 per cent one magnitude fainter. With the
fiducial PSF/blue-SED response held fixed, simulations detect these eight
cases in only 5--26 per cent of trials, using null-calibrated detection
thresholds with a 5 per cent false-positive rate, so the test sets no turnover
exclusion limits. The observed counts do not require
extra curvature beyond an unrestricted Schechter function. A steep, assumed
30-per-cent B6 contaminant template can produce a marginal preference
for curvature. The result is conditional on
the adopted source profiles, lens reconstruction and negligible residual
contamination, and applies to this single sightline.}
\end{abstract}

\section{Introduction}
\label{sec:introduction}

The rest-frame ultraviolet luminosity function (UVLF) traces unobscured star
formation and connects the abundance of early galaxies to cosmic reionization.
Blank-field surveys establish its broad evolution over $z\simeq2$--10
\citep{McLure2013,Bouwens2015,Finkelstein2015,Bouwens2021,Harikane2022},
while JWST has extended the accessible redshift range and exposed sensitivity
to photometric selection and model assumptions
\citep{Donnan2023,Harikane2023,Bouwens2023Selection}.

Strong-lensing clusters probe intrinsically fainter galaxies, but magnification
changes both observed flux and source-plane area.  Multiple images must be
associated with one physical source, and completeness changes rapidly near the
detection limit.  Hubble Frontier Fields analyses demonstrated both the reach
and the lens-, size-, and selection-dependence of this approach
\citep{Lotz2017,Shipley2018,Atek2015,AtekCombined2015,Castellano2016,
Livermore2017,Bouwens2017,Ishigaki2018,Kawamata2018,Atek2018,
BouwensSize2022,Bouwens2022a,Bouwens2022b}.  A decrease in raw faint counts is
therefore not, by itself, evidence for a physical turnover.

Abell~2744 is particularly useful for a source-plane reassessment.  UNCOVER
provides deep HST and JWST imaging, multiband catalogs, NIRSpec spectroscopy,
and public strong-lensing products \citep{Bezanson2024,Weaver2024,Suess2024,
Furtak2023,Price2025,UNCOVERDR4}.  The field has also supported complementary
population studies with explicit selection-response calculations
\citep{Chemerynska2024,Wold2026}.  We analyze the total UV-selected population
at $6\leq z<8$, rather than an emission-line or $z>9$ subsample.

An earlier preprint interpreted the same field as evidence for global
faint-end suppression \citep{Ma2025}.  The present paper supersedes that
interpretation and asks the narrower model-selection question: within the
range supported by \revised{the calibrated response}, do the data require curvature in
addition to an unrestricted Schechter function? \rtwo{To answer it, we build
a forward model that connects a physical LF to fixed-aperture counts,
including counterimages, and calibrate it with artificial sources added to
the images (section~\ref{sec:analysis}). Section~\ref{sec:results} reports
the depth of this calibration, the sensitivity of the test to curvature and
a comparison with GLIMPSE \citep{Atek2026}, and section~\ref{sec:discussion}
discusses the assumptions behind the result.}
\revised{Luminosity distances and comoving
volumes use the Planck 2018 cosmology \citep{Planck2020}.  We project the
fixed public lens reconstructions using their original flat
$\Lambda$CDM distance convention, $H_0=70\,\mathrm{km\,s^{-1}\,Mpc^{-1}}$
and $\Omega_{\rm m}=0.3$, with lens redshift 0.308. The mass models are not
refitted for a different cosmology.}  Magnitudes are on the AB system.

\section{Data and sample}
\label{sec:data}

\subsection{Imaging and photometric conventions}
\label{sec:photometry}

We use public UNCOVER DR3 mosaics and the D032 catalog
\citep{Weaver2024,Suess2024,UNCOVERDR3Data}.  The observed selector reads the
released \texttt{f\_fXXXw} and \texttt{e\_fXXXw} columns.  These are derived
from 0.32-arcsec-diameter color apertures on PSF-matched images and include the
catalog's source-dependent, band-common aperture-to-total factor
\texttt{tot\_cor}, so they are not raw aperture fluxes.
\revised{We retain the catalog's Galactic-extinction correction
\citep{FitzpatrickMassa2007} in the flux and error conventions.}
\revised{F277W, F356W, and F444W form the detection stack.}  Selection uses F814W,
F115W, F150W, F200W, F277W, F356W, and F444W fluxes and uncertainties.
\revised{Table~\ref{tab:imaging} lists the published median depths. We use
each source's flux uncertainties in the redshift fit. To measure the
selection, we add artificial sources, synthetic galaxies of known brightness
and redshift, to the real images and apply the same photometric-redshift fit
and selection to them. Their recovery rates predict how many galaxies enter
each observed luminosity bin (subsections~\ref{sec:injections}
and~\ref{sec:likelihood}).}

\begin{table*}[t]
\tbl{The seven imaging bands and their published median 5$\sigma$ depths in
0.32-arcsec-diameter apertures.}{%
\begin{tabular}{llcl}
\toprule
Band & Instrument & Depth (AB) & Analysis role \\
\midrule
F814W & HST/ACS & 27.58 & \revised{optical non-detection and SED fit} \\
F115W & JWST/NIRCam & 29.18 & \revised{S/N selection and SED fit} \\
F150W & JWST/NIRCam & 29.14 & \revised{S/N selection and SED fit} \\
F200W & JWST/NIRCam & 29.23 & \revised{S/N selection and SED fit} \\
F277W & JWST/NIRCam & 29.57 & \revised{detection stack and SED fit} \\
F356W & JWST/NIRCam & 29.70 & \revised{detection stack and SED fit} \\
F444W & JWST/NIRCam & 29.21 & \revised{detection stack, PSF target, and SED fit} \\
\bottomrule
\end{tabular}}
\label{tab:imaging}
\end{table*}

\revised{We require \texttt{use\_phot}=1 and reject released
flags for missing photometry, low detection S/N, stars, artifacts or
proximity to a subtracted bright cluster galaxy (BCG).
The released \texttt{flag\_eazy} describes the catalog's original quicklook
fit. We instead require a finite, successful new posterior fit and the
selection below, and test the old EAZY-flag requirement separately.
Where an associated DR4.1 spectrum has quality $\geq2$ and a
successful finite best-fit redshift outside $6\leq z<8$, we exclude that
specific source \citep{UNCOVERDR4}.  A quality-$\geq3$ version and the
photometry-only sample are separate checks.}

\revised{Selection requires $p_{68}=P(6\leq z<8)\geq0.60$ and
$p_{68}/P(z<6)\geq3$, together with either of two S/N routes:
F115W/F150W $\geq5/3$, or F150W/F200W $\geq5/3$.
Each route requires valid measurements in its two near-infrared bands and
F814W, with signed F814W S/N $\leq2$.  A valid measurement has finite flux, positive finite error
and positive finite catalog weight. Negative
F814W flux is allowed.  Passing either route admits the object, after which
the larger posterior mass assigns B6 ($p_6\geq p_7$) or B7 ($p_7>p_6$),
where $p_6=P(6\leq z<7)$ and $p_7=P(7\leq z<8)$.
Thus the S/N route and assigned redshift bin need not agree.}
\revised{The published errors include the empirical background-noise
calibration. We add a 5 per cent flux-proportional floor in quadrature
without rescaling the correlated noise a second time.
Artificial-source colors use the released F444W PSF-matching kernels,
0.32-arcsec forced apertures, and a band-common correction for each prescribed
image profile. Their errors follow the released aperture-noise versus local
weight relation. Initial detection uses the native, noise-equalized long-wave
stack, whereas color photometry uses PSF-matched images. The real and artificial
sources share the EAZY configuration
\citep{Brammer2008}, Inoue intergalactic attenuation \citep{Inoue2014}, and
selector. The Supplementary Data gives the settings and local measurement
checks.}

\revised{Before fitting the posteriors, we screen the catalog with the two
S/N routes and the optical veto. This saves computation without adding a
selection cut.}
\revised{Of 74,020 D032 entries, 8450 pass the preliminary screen, 261 the
redshift selector, and 242 the retained quality requirements. Imaging
coverage retains 227 of these, finite lens support 190, and $0<\mu\leq30$
189. Removing one registered duplicate leaves 188 sources,
all within the clean region below. Of these, 14 have an associated
quality-$\geq2$ DR4.1 spectrum. Four of the 14 lie outside the target
interval: three B6 sources at $z_{\rm spec}=5.67$--5.99, just below the B6
boundary, and one B7 source at 1.42 whose spectrum has quality 2. Excluding
the four leaves 184 sources (124 B6, 60 B7), before the fitted-magnitude
cut. Retaining the quality-2 object instead gives 185 (124 B6, 61 B7).}

\subsection{Footprint, lens products, and source identity}
\label{sec:footprint}

\begin{figure*}[t]
\begin{center}
\includegraphics[width=0.94\linewidth]{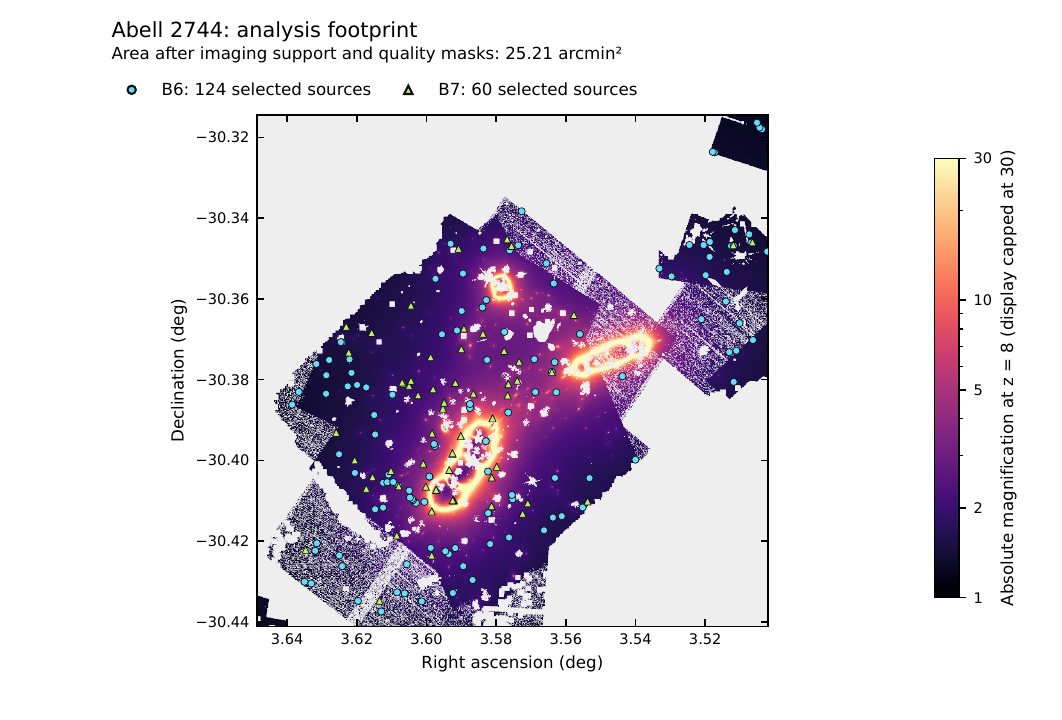}
\end{center}
\caption{\revised{The analysis footprint in Abell~2744 after imaging-support
and quality masks.  Color shows the absolute UNCOVER v2.0 best-fit
magnification at $z_{\rm s}=8$, capped at 30 for display. Grey regions are
outside the supported clean domain.  Cyan circles and green triangles mark
the 124 B6 and 60 B7 registry-de-duplicated source representatives before
the fitted-magnitude range is imposed.  The supported image-plane area is
$25.21\,\mathrm{arcmin}^{2}$.}\\
{\revised{Alt text: A magnification map of the Abell 2744 analysis region, with grey
excluded areas, cyan circles for B6 sources and green triangles for B7 sources.}}}
\label{fig:footprint}
\end{figure*}

The geometric footprint is the intersection of full square support in the
three detection mosaics, evaluated with a 56-pixel margin, and finite support
in all four UNCOVER v2.0 lens maps
\citep{Furtak2023,Price2025,UNCOVERDR4,UNCOVERLensData}.
Sampled on the lens grid, the intersection covers
$33.36\,\mathrm{arcmin}^{2}$.
\revised{Within this geometric frame, the analysis requires native-weight
support in F814W, F115W, F150W and F200W. We define a clean region using
public segmentation labels whose catalog entries carry a
stellar, artifact or near-BCG flag.  Their pixels are dilated by an integer
disk of radius four native 40-mas pixels, corresponding to the 0.16-arcsec
color-aperture radius.  The remaining supported image-plane area is
$25.21\,\mathrm{arcmin}^{2}$. We built this mask for our analysis
because the original star-spike and BCG masks are unavailable. We retain other
galaxy segments, so the selected region includes occupied as well as blank
sky. The same spatial rule applies to observed positions, injection truth,
recovered positions and lens-branch support before source-plane integration.
Because the catalog flags cannot be assigned to artificial sources, this mask
applies the same rejection to both. Among the 261 real selector candidates,
it admits exactly the 242 that pass the released flags
(Supplementary Data).}
Figure~\ref{fig:footprint} shows the domain and \revised{the selected source representatives}.

\revised{The public UNCOVER v2.0 best-fit map defines the fiducial analysis.}
\rtwo{Read with the documented normalization, the 50 released model-range
maps are larger than the best-fit model by a uniform factor of about 1.67.
We confirmed with the UNCOVER lens modellers that a pixel-scale conversion
was omitted when these maps were computed. Scaled by 0.60, they reproduce the release's own magnification
uncertainties, and they then give a statistical magnification uncertainty of
0.01\,mag (median) for our sources (Supplementary Data).}
\revised{Our fiducial result is conditional on the best-fit reconstruction. Variation within one model
family and differences between independent models probe different uncertainties}
\citep{SchneiderSluse2013,Priewe2017,Meneghetti2017,Raney2020,
Bergamini2023}.  Earlier HST and MUSE constraints establish the model lineage
\citep{Jauzac2015,Mahler2018}.

A public multiple-image registry is used to associate known counterimages.
\revised{One representative per registered source enters the observed counts},
while all valid counterimage opportunities remain in the \revised{counterimage
union of equation~(\ref{eq:union})}.  We therefore describe
the catalog as registry-de-duplicated. \revised{The coarse lens mapping
predicts multiple covered image regions for 18 of the 184 representatives.
These model-predicted candidates are concentrated among the most magnified
sources: 15 of the 21 sources with $\mu>5$ are candidates. All 15 lie
faintward of the fitted range. Only two of the 151 fitted entries are
candidates, and the fitted sample has median magnification 1.86. Separately, we search all 184
representatives for close source-plane pairs and finite counterimage
regions using their redshift posteriors. This search finds eighteen pairs
within 0.6 arcsec at either their individual or common mean redshift,
which we check against colors, parity and available image/segmentation
evidence.
Close source-plane positions alone do not establish an association.
19308/19309/19310 occupy one local image structure, but the data do not
resolve whether these entries are galaxy knots or distinct galaxies.
Entry 24575 remains a possible distant counterimage. We therefore retain the
registry associations and test alternative counts without summing potentially
blended fluxes. The Supplementary Data gives the spectroscopic association
checks, spectral matching and quality criteria}
\citep{Price2025,UNCOVERSpectroscopyData}.

\section{Analysis}
\label{sec:analysis}

\revised{We select galaxies by color and redshift posterior, associate known
counterimages, and measure an aperture-based luminosity proxy. We add
artificial sources to the images to measure, for each true luminosity and
redshift, how often a galaxy is selected, which redshift interval it is
assigned to and what proxy magnitude is recovered. We call this measured
mapping the response (subsections~\ref{sec:injections}
and~\ref{sec:likelihood}). The lens mapping and comoving volume then give predicted
counts for a trial LF, counting each source-plane cell once. Artificial
sources do not enter the galaxy sample.}

\subsection{\revised{True luminosity and the observed aperture proxy}}
\label{sec:m1500}

At each redshift node the EAZY best-fitting SED is integrated through a
1450--1550~\AA\ rest-frame top-hat. \revised{The catalog-total magnitude estimate is}
\begin{equation}
\revised{M_{1500}^{\rm cat}=m_{1500(1+z)}^{\rm cat}-\mathrm{DM}(z)
 +2.5\log_{10}(1+z)+2.5\log_{10}\mu,}
\label{eq:m1500}
\end{equation}
\revised{The $2.5\log_{10}(1+z)$ term follows the AB flux-density
K-correction convention \citep{Hogg2002}.} The last term removes lens
magnification.  For each selected object we
place eight equal-posterior-mass midpoint nodes within its assigned B6 or B7
interval, refit the SED at each fixed redshift, and adopt the median
\revised{$M_{1500}^{\rm cat}$}. The fiducial magnification is evaluated at the interval-conditional
mean redshift. Thus the rest-frame K-correction and redshift dependence are
\revised{retained}, while hard B6/B7 membership remains the sample definition.

\revised{We construct a common fixed-aperture observable for the likelihood
to avoid the catalog's noisy Kron total correction. The release applies one
positive total correction $q_{\rm cat}$ to all seven aperture fluxes and
errors of each object. We replace it by the fixed F444W PSF correction
$q_0\simeq1.55017$, giving}
\begin{equation}
\revised{\widetilde M_{1500}=M_{1500}^{\rm cat}
                 +2.5\log_{10}(q_{\rm cat}/q_0).}
\label{eq:apertureproxy}
\end{equation}
\revised{This scalar change preserves valid bands, colors, S/N and the
no-prior redshift likelihood at every conditional node. All artificial
profiles use the same $q_0$. The observable $\widetilde M$ measures aperture
light, whereas the LF remains defined in true total $M$. The response from $M$ to
$\widetilde M$, measured with the artificial sources, includes
profile-dependent aperture loss, noise and redshift errors. Observed and
injected sources are therefore compared on the same scale without
reconstructing the catalog's Kron and deblending procedure.}

\revised{The same rest-filter definition maps each injected SED to apparent
F150W magnitude at the eight effective-volume nodes, four in each interval.
We update all seven predicted colors with redshift, including the six
near-infrared bands as well as F814W. The two stellar-population SEDs are
members of the fitting library, and each defines one calibration scenario.}

\subsection{Joint detection and seven-band selection response}
\label{sec:injections}

\revised{We draw 256 independent background positions without replacement
from 36 strata combining foreground environment and F814W/F150W depth,
retaining positions occupied by existing segments.  Every position receives
eight redshifts, 17 apparent F150W magnitudes from 19 to 39, two redshifted
stellar-population SEDs (nominal $A_V=0.01$, `blue', and 0.25, `red') and
five prescribed image profiles. \rtwo{These are the PSF and the PSF
convolved with circular Gaussians of $\sigma=0.05$ (the broadened PSF),
0.10, 0.20 and 0.40 arcsec. The primary analysis uses the two most compact
profiles, the PSF and the broadened PSF, each with both SEDs. These form its
four profile/SED calibrations, and the primary results are quoted as ranges
over them. The PSF with the blue SED is the fiducial response for the tests
that are too costly to repeat for every calibration: the power simulations,
null refits, calibration resamples and GLAFIC comparison. The
$\sigma=0.10$-arcsec profile tests a larger size with the full forward
response (subsection~\ref{sec:sensitivities}). The 0.20 and 0.40-arcsec
profiles reach only $-18.0$ to $-19.5$ in the five-per-cent volume
diagnostic, 1.5--3\,mag brighter than the PSF. They therefore cannot test
the faint end (Supplementary Data).}
The 348,160 visits thus reuse 256 backgrounds and cannot be treated as
independent noise realizations. For each visit, we add the signal to the
untouched seven-band background, run long-wave detection, measure colors at
the detected centroid and apply the common redshift selector. Artificial
sources cannot carry catalog flags, so we apply the cut behind the low-S/N
flag directly: a signed S/N of at least 3 in the PSF-matched
F277W+F356W+F444W aperture. On the released photometry this cut reproduces
the flag for all 74,020 catalog entries (Supplementary Data). Failed visits
remain in the recovery denominator. The adopted clean domain retains 251
backgrounds and 341,360 visits for the final response.}

\begin{figure*}[t]
\begin{center}
\includegraphics[width=0.98\linewidth]{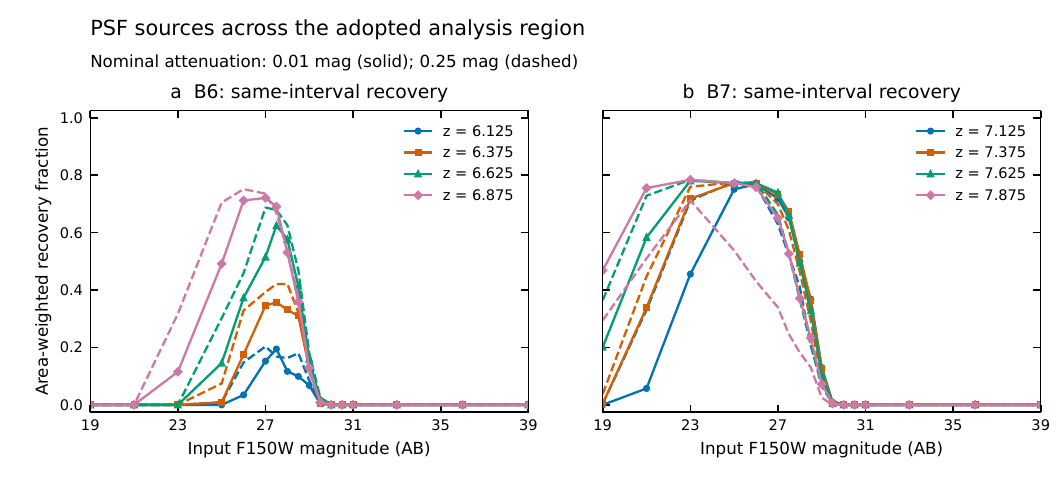}
\end{center}
\caption{\revised{Measured same-interval recovery for artificial PSF sources
in the adopted analysis region.  Color and marker shape identify input
redshift. Solid and dashed curves use \rtwo{stellar-population} SEDs with nominal $A_V=0.01$
and 0.25. Inverse inclusion probabilities weight the 251 of 256 sampled
background positions inside the clean domain. Lines connect measured magnitude nodes without a
monotonic constraint.  The bright-end decline includes rejection by the
optical non-detection cut.  These spatially averaged curves illustrate the
selection. The effective volumes use the local depth and environment
response at each lens branch.}\\
{\revised{Alt text: B6 and B7 recovery fractions rise from the bright end and then
fall toward faint magnitudes.  Four redshift colors and two line styles
show stronger redshift variation for B6.}}}
\label{fig:selector-response}
\end{figure*}

\revised{Figure~\ref{fig:selector-response} illustrates the measured recovery.
We estimate it locally as a function of observing depth and
foreground environment, weighted by the recorded inverse inclusion
probabilities. To choose predictors and neighbourhood size, we hold out
entire backgrounds, excluding all their repeated conditions together.
We interpolate only within the measured brightness grid and allow
non-monotone response curves, because even a securely detected bright source
can fail the optical veto. We record recovery in the input redshift interval
and transfer to the other interval separately. We evaluate each SED/profile case on its own, because assigning equal
weights would impose a galaxy size distribution that we have not measured
\citep{Atek2018,BouwensSize2022}.  Completeness and flux-migration conventions
are discussed by \citet{Leethochawalit2022}.}

\subsection{Branch-aware source-plane effective volume}
\label{sec:veff}

The lens maps are sampled on a stride-four image grid and mapped to a fixed
source grid at four redshift nodes in each interval. \revised{The native
lens-map pixel is 0.1 arcsec and the source-cell side is 0.4 arcsec.}
\revised{This accounts for} the source-plane area required for magnification calculations
\citep{VegaFerrero2019}.  For intrinsic magnitude $M$ and redshift $z$, the
lensed apparent magnitude is
\begin{equation}
\begin{split}
m_i(M,z)={}&M+\mathrm{DM}(z)-2.5\log_{10}(1+z)\\
&-2.5\log_{10}\mu_i(z)+\Delta_{\rm SED}(z),
\end{split}
\end{equation}
\revised{Here $m_i$ is the apparent F150W magnitude of image $i$,
$\mathrm{DM}$ is the distance modulus, $\mu_i$ is the absolute magnification,
and $\Delta_{\rm SED}$ maps the injected colors to the rest-frame definition.
Connected rays of equal image parity that map to one source cell form one
image branch and count as one detection opportunity. For morphology $p$,
the branch-specific detection-and-selection probability is $C_{b,p}$.
Distinct counterimage opportunities of source cell $s$ are united as}
\begin{equation}
P_{s,p}(M,z)=1-\prod_{b\in\mathcal B_s}[1-C_{b,p}(M,z)].
\label{eq:union}
\end{equation}
\revised{This expression assumes that recovery in separated images is
independent at fixed signal, redshift, SED and morphology. Cells with more
than one active image contribute 1.9 per cent of the expected fitted counts
under the fiducial response.
\rtwo{Each profile/SED calibration assumes that all galaxies share that
profile and SED. The response of a population with a mixture of sizes and
SEDs is the weighted mean of these responses, taken after the image union
because all images of a source share one profile and SED. We do not measure
the weights, so we report each calibration separately and test equal-weight
mixtures (subsection~\ref{sec:sensitivities}).} We sum over
source-cell solid angle to obtain $\Omega_{\rm eff}$ in steradians, then
over comoving shells:}
\begin{equation}
V_{\rm eff}(M)=\sum_j\Omega_{\rm eff}(M,z_j)
\frac{V_{\rm c}(z_{j,+})-V_{\rm c}(z_{j,-})}{4\pi}.
\label{eq:veff}
\end{equation}
\revised{$V_{\rm c}$ denotes the full-sky comoving volume, so division by
$4\pi$ converts each shell to volume per steradian.  Counterimages can improve
detectability without multiplying physical source-plane area, so no additional
$1/\mu$ factor is applied after source-plane cells have been counted.
The $0<\mu\leq30$ cut is applied at each outcome's recovered,
interval-conditional mean redshift before choosing the representative
image. Thus outcomes may enter or leave this cut relative to the true-redshift
magnification. The branch topology is built before the magnification cap is
applied.}

\Needspace{4\baselineskip}
\subsection{\revised{Forward response and count likelihood}}
\label{sec:likelihood}

\revised{At each true luminosity and redshift, the empirical response jointly
records selection, assigned redshift, $\widetilde M$ and F150W S/N.
For multiply imaged sources, the highest-F150W-S/N selected image supplies
the observed bin, as in the catalog. Integrating this representative's joint
distribution preserves the dependence of magnitude error on S/N.
Summing over accepted categories recovers
equation~(\ref{eq:union}).}

\revised{Let $P_k(s,M,z_j)$ be the probability that source cell $s$ is
recorded in observed redshift/proxy-magnitude bin $k$. For a true-redshift
interval $t$, define the response volume}
\begin{equation}
\revised{K_{kt}(M)=\sum_{j\in t}\frac{\Delta V_{{\rm c},j}}{4\pi}
                         \sum_s\Delta\Omega_s\,P_k(s,M,z_j).}
\label{eq:responsekernel}
\end{equation}
\revised{It has units of $\mathrm{Mpc}^3$. Summing $K_{kt}$ over all accepted
output categories gives the contribution of true interval $t$ to the
effective volume of equation~(\ref{eq:veff}). The likelihood uses only the
bin-resolved $K_{kt}$. The expected count is then}
\begin{equation}
\revised{\lambda_k(\boldsymbol\vartheta)=
 \sum_{t\in\{\mathrm{B6,B7}\}}\int_{\mathcal M}
 \phi_t(M\mid\boldsymbol\vartheta_t)K_{kt}(M)\,dM.}
\label{eq:binintegral}
\end{equation}
\revised{The integral covers the calibrated true-total-luminosity domain
$\mathcal M$, rather than the observed bin $k$, to retain migration in
magnitude and between the two redshift intervals. We integrate piecewise across response-grid
knots and keep zero-count bins in the fitted observation range. Exact central
68.27 per cent Poisson intervals describe the displayed observed counts
\citep{Garwood1936}. Because galaxies migrate between proxy bins, which a
diagonal volume such as $V_{\rm eff}$ ignores, we compare the physical model
with the data through predicted counts.}

\revised{This baseline includes migration within $6\leq z<8$ and assumes
negligible residual contamination from outside that interval. The
Supplementary Data tests specified additive contaminant-count templates.}

\revised{The response is evaluated on 101 true-total-magnitude nodes
from $-23$ to $-13$, retaining that complete domain in every fit.
For each recovered redshift interval we retain half-magnitude observation
bins whose unit-density exposure,
$E_k=\sum_t\int_{-23}^{-13}K_{kt}(M)\,dM$, exceeds 5 per cent of its
interval maximum in all four compact-profile/SED scenarios. This common
mask is determined by the response alone, without using the observed counts.
The threshold measures relative exposure, not the detection probability of
an individual source.
Separately, the true-luminosity range supported by those observations is
where $\sum_{k\,\mathrm{retained}}K_{kt}(M)$ exceeds 5 per cent of its
maximum. This range limits where we plot and interpret the physical LF, but
it does not limit the integration in equation~(\ref{eq:binintegral}).}

The null model is the unrestricted Schechter function \citep{Schechter1976},
\begin{equation}
\begin{split}
\phi_{\rm Sch}(M)={}&0.4\ln 10\,\phi_*\,
10^{-0.4(M-M_*)(\alpha+1)}\\
&\times\exp[-10^{-0.4(M-M_*)}],
\end{split}
\label{eq:schechter}
\end{equation}
and the alternative multiplies it by
\begin{equation}
S(M\mid M_{\rm t},\beta)=
\begin{cases}
1, & M\leq M_{\rm t},\\
10^{-0.4\beta(M-M_{\rm t})^2}, & M>M_{\rm t},
\end{cases}
\label{eq:curvature}
\end{equation}
with \revised{$0\leq\beta\leq4$ and $-21.5\leq M_{\rm t}\leq-13$}.  The null is recovered at $\beta=0$, and shared parameters
have identical bounds.  We maximize the independent Poisson likelihood
\citep{Cash1979} and define
$\Delta\mathrm{IC}=\mathrm{IC}_{\rm curved}-\mathrm{IC}_{\rm Sch}$. Positive
values favor the simpler model after the AIC or BIC penalty
\citep{Akaike1974,Schwarz1978}.
\revised{We use $\mathrm{AIC}=2k-2\ln\mathcal L_{\max}$ and
$\mathrm{BIC}=k\ln n-2\ln\mathcal L_{\max}$, where the joint B6/B7 null
has $k=6$. Adding curvature in either true-redshift interval gives $k=8$, and
the other interval's three baseline parameters are refitted simultaneously.
Here $n$ counts the fitted Poisson observation bins, including zero-count bins.
Writing
$T=2(\ln\mathcal L_{\rm curved}-\ln\mathcal L_{\rm Sch})$ gives
$\Delta\mathrm{AIC}=4-T$.}
Since $M_{\rm t}$
is unidentified at $\beta=0$, these are descriptive model comparisons, not
likelihood-ratio significances.  \revised{Count noise and finite calibration
are separate uncertainty sources. Calibration uses 32 stratified
whole-background resamples of the fiducial PSF/blue-SED response
\citep{Efron1979}. Each resample keeps the repeated conditions of a
background together.
For count noise, 256 Poisson realizations of the PSF/blue-SED null are
jointly refitted at fixed response. Their displayed 16th--84th percentiles
are pointwise refit distributions that exclude calibration uncertainty
(Supplementary Data).}

\section{Results}
\label{sec:results}

\subsection{\revised{Sample and luminosity response}}
\label{sec:sampleresponse}

\begin{figure*}[t]
\begin{center}
\includegraphics[width=0.98\linewidth]{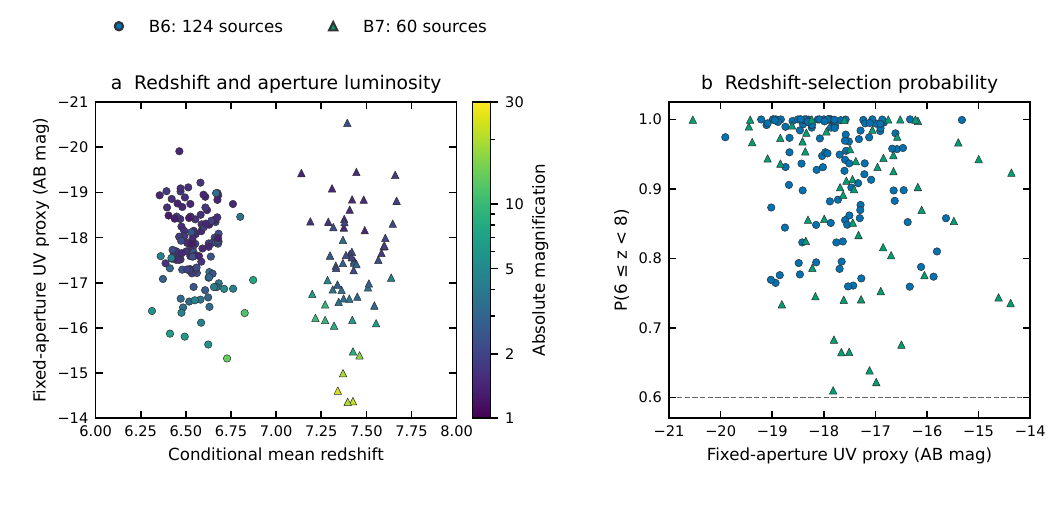}
\end{center}
\caption{\revised{Source distribution before fitted-magnitude
selection.  The left panel shows assigned-interval conditional mean
redshift and the fixed-aperture proxy $\widetilde M_{1500}$, with color indicating
magnification.  Circles denote B6 and triangles B7.  The right panel shows
target-interval posterior mass against $\widetilde M_{1500}$, and the dashed
horizontal line marks the $p_{68}=0.6$ selection threshold. Both panels use aperture
light rather than total-galaxy luminosity.}\\
{\revised{Alt text: Two scatter plots. The left plots redshift against the
demagnified aperture-luminosity proxy, with points colored by magnification.
The right plots posterior membership probability against that proxy with a horizontal
selection-threshold line.}}}
\label{fig:sample}
\end{figure*}

\revised{The 184 registry-de-duplicated sources have mean
target-interval posterior mass 0.915.  Figure~\ref{fig:sample} shows all
124 B6 and 60 B7 sources before imposing a fitted-magnitude range.
Their demagnified aperture proxies $\widetilde M$ span $-19.91$ to $-15.32$
in B6 and $-20.54$ to $-14.36$ in B7. These extrema describe the measured
sample. The limits in true total magnitude depend on the selection and
luminosity-response model.}

\begin{figure*}[t]
\begin{center}
\includegraphics[width=0.98\linewidth]{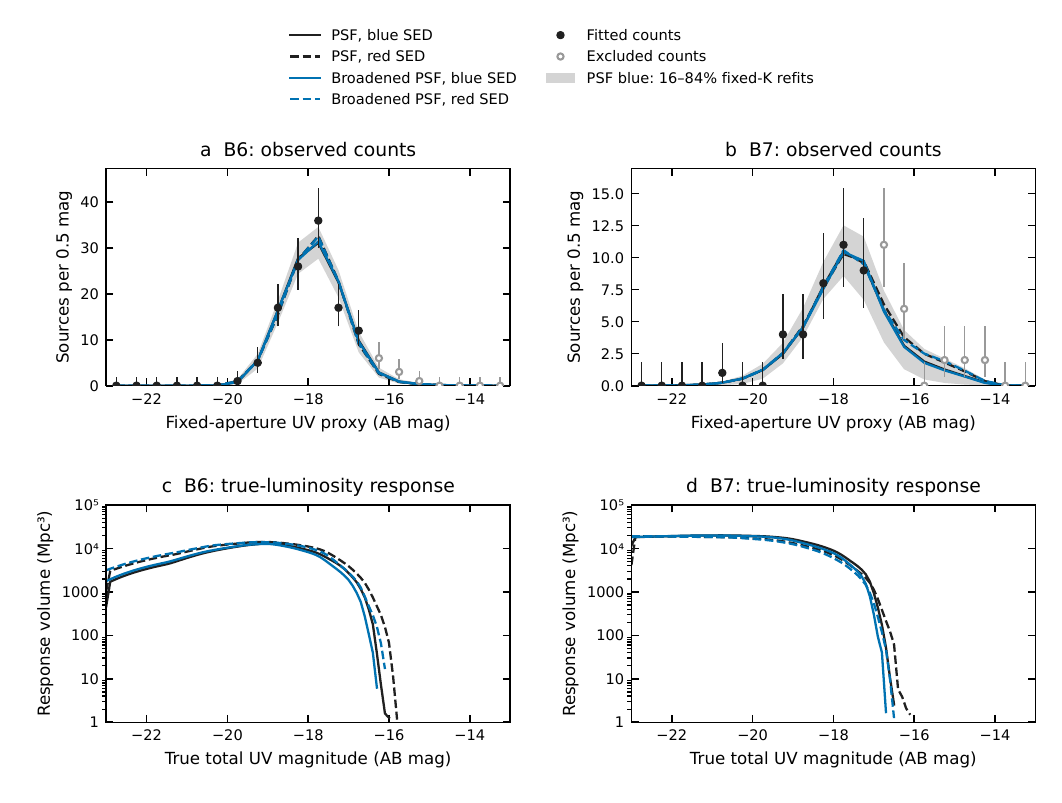}
\end{center}
\caption{\revised{Observed counts and the forward response. Upper panels
show counts in fixed-aperture proxy bins, with exact central 68.27 per cent
Poisson intervals. Filled points enter the common fit and grey open points
do not. The first excluded faint bins lie above the predictions
(subsection~\ref{sec:sampleresponse}). Each curve is a joint Schechter prediction under one compact
profile/SED assumption, so the four curves do not form an uncertainty
envelope.
Grey shading gives
the 16th--84th percentiles of predicted counts from 256 fixed-response
PSF/blue-SED null refits and excludes calibration uncertainty. Lower panels show
the response volume into retained observations as a function of true total
magnitude.}\\
{\revised{Alt text: Four panels. Upper panels compare observed proxy-magnitude
counts and four forward predictions. Lower panels show the selection
volume versus true total magnitude for each conditional calibration.}}}
\label{fig:veff-uvlf}
\end{figure*}

\revised{The common observation mask retains 13 B6 bins over
$-23\leq\widetilde M<-16.5$ and 12 B7 bins over
$-23\leq\widetilde M<-17$, containing 114 and 37 sources, respectively.
Figure~\ref{fig:veff-uvlf} compares their counts directly with forward
predictions. Across the four scenarios, the faint true-total-magnitude
five-per-cent relative-peak boundaries range from $-16.71$ to $-16.39$ in
B6 and $-17.13$ to $-16.96$ in B7. The corresponding 50-per-cent boundaries
are $-17.71$ to $-17.41$ and $-18.53$ to $-18.14$, respectively.
These describe the response into retained observations, not individual-source
completeness or the observed proxy-bin edges. The Supplementary Data gives
each boundary and tests the integration endpoints.}

\revised{Immediately outside the fitted faint edges, B6 has six sources in
$[-16.5,-16)$ versus 2.56--2.88 predicted by the fitted nulls, and B7 has
11 in $[-17,-16.5)$ versus 5.79--6.35. Both bins lie outside the fitted range.}

\subsection{Model comparison and core sensitivities}
\label{sec:sensitivities}

\revised{Table~\ref{tab:modelcomparison} compares the joint fits to the
same observations. All four B6 curved alternatives return $\beta=0$ with
no likelihood gain. In B7, the improvement is only $T=0.050$--0.134,
far less than the two-parameter AIC penalty of four. With 25 fitted bins,
$\Delta\mathrm{BIC}=+6.44$ for B6 and $+6.30$--$+6.39$ for B7.
All four null optima lie inside the parameter bounds, but three B7 curved
optima reach the bright $M_*=-24$ boundary, leaving the alternative's shape
parameters poorly constrained. The Supplementary Data gives the null
parameter estimates, common bounds and supporting sensitivity tests.
For fiducial B7, the 256 fixed-response null refits give a 16th--84th
percentile range $\alpha=-2.77$ to $-1.73$. This spread includes fits at the $M_*$ bounds.
All 32 calibration resamples of the fiducial response retain positive
$\Delta\mathrm{AIC}$ for both alternatives.}

\revised{Simulations at the fixed fiducial PSF/blue-SED response detect the
eight tested curvature scenarios only 5--26 per cent of the time. Each
scenario combines B6 or B7, $M_{\rm t}=-18.5$ or $-17.5$, and $\beta=0.25$
or 1, with 128 Poisson realizations and all six baseline parameters
refitted. These factors suppress density by 21 or 60 per cent one magnitude
beyond the onset, but a free Schechter shape absorbs much of this change
over the accessible range. Separate detection thresholds, each set for a
5 per cent false-positive rate, come from 256 null realizations
(Supplementary Data). The observed fiducial values, $T=0$ in B6 and 0.07 in
B7, lie below these thresholds ($T>3.84$ and $T>1.48$). The
$\Delta\mathrm{AIC}<0$ rule ($T>4$) is stricter. On the same simulations,
it detects the tested curvature in 4--20 per cent of B6 trials and 0--5 per
cent of B7 trials. The null fit itself describes the data adequately: about
59 per cent of the refitted null simulations have a deviance at least as
large as the observed one. For the fiducial response, the summed response
into the retained bins is zero for true $M\geq-15.9$ in B6 and
$M\geq-16.2$ in B7. Onsets at or fainter than these magnitudes therefore
cannot change the retained-bin predictions. These zero-response limits lie
faintward of the five-per-cent boundaries of
subsection~\ref{sec:sampleresponse} and are not physical depth limits.}

\begin{table*}[t]
\begingroup
\tbl{\revised{Joint forward fits to the common 25 observation bins (114 B6
and 37 B7 sources), one row per profile/SED calibration. Slopes are
null-model point estimates. Positive $\Delta$AIC favors retaining the null.
SED labels denote the two injected templates.}}{%
\begin{tabular}{llrrrr}
\toprule
Profile & SED & $\alpha_{\rm B6}$ & $\alpha_{\rm B7}$ & $\Delta$AIC (B6) & $\Delta$AIC (B7) \\
\midrule
PSF & blue & $-0.869$ & $-2.582$ & $+4.00$ & $+3.93$ \\
PSF & red & $-0.939$ & $-2.706$ & $+4.00$ & $+3.89$ \\
Broadened PSF & blue & $-0.978$ & $-2.549$ & $+4.00$ & $+3.87$ \\
Broadened PSF & red & $-1.026$ & $-2.740$ & $+4.00$ & $+3.95$ \\
\bottomrule
\end{tabular}}
\label{tab:modelcomparison}
\endgroup
\end{table*}

\rtwo{To test a larger image profile, we reconstruct the complete forward
response, including selection and magnitude/redshift migration, for a PSF
convolved with a $\sigma=0.10$-arcsec Gaussian from the existing seven-band
trials. On the same 151 counts, the two SED cases give B6
$\Delta\mathrm{AIC}=4.00$ and B7 3.87--3.93. Equal-weight mixtures of the
four primary calibrations, and of these with the two $\sigma=0.10$-arcsec
cases, give B6 $\Delta\mathrm{AIC}=4.00$ and B7 3.92. The ranking is
therefore unchanged for a larger profile and for mixed populations of these
profiles and SEDs.}

\revised{Contamination sensitivity depends on its luminosity distribution.
An additive B6 contaminant-to-baseline ratio rising by a factor 2.51 per
magnitude retains positive $\Delta\mathrm{AIC}$ at tested fractions up to
60 per cent of the baseline B6 counts. A steeper factor of 6.31 per
magnitude at 30 per cent gives B6 $\Delta\mathrm{AIC}=-0.05$ to $+1.16$.
Increasing the curvature bound from $\beta\leq4$ to 16 changes this to
$-0.58$ to $+0.82$. These weak, bound-dependent preferences qualify the
baseline result (Supplementary Data). The data do not show that residual
contamination is negligible. Three of the four spectroscopic exclusions lie
just below $z=6$ (subsection~\ref{sec:photometry}), and the mean target-interval
posterior mass of 0.915 means that about 16 of the 184 sources are expected
to lie outside $6\leq z<8$. Neither number is a calibrated contamination
rate.}

\revised{An independent calculation with published GLAFIC maps
\citep{Oguri2010,Kawamata2016,Kawamata2018,GLAFICv4Data} covers the same clean sky region
as the corresponding UNCOVER calculation, $7.63\,\mathrm{arcmin}^{2}$,
with model-specific magnification cuts. This region, limited by the extent of
the public GLAFIC map, is about 30 per cent of the fiducial image-plane area.
For the fiducial PSF/blue-SED calibration the two responses define a common
mask, 0.5\,mag deeper than the full-field mask, that retains 64 UNCOVER and
65 GLAFIC sources. Both give $T\simeq0$ and $\Delta\mathrm{AIC}=+4.00$
for either curvature alternative, and both B7 nulls reach $M_*=-24$.
This restricted-area test leaves the ranking unchanged (Supplementary Data).}

\revised{Counting 19308/19309/19310 once, with or without associating
24575, changes only the B7 proxy bins centred at $-16.25$ and $-16.75$.
Both lie outside the common fitting interval, leaving the 151 primary
counts unchanged. Other unresolved local pairs include fitted entries.
A separate count-only test keeps one representative for each of six such
pairs. This leaves 146 fitted entries and gives
$\Delta\mathrm{AIC}=+3.70$--$+4.00$ across the four calibrations and both
alternatives (Supplementary Data).}

\subsection{Comparison with published UVLFs}
\label{sec:comparison}

\begin{figure*}[t]
\begin{center}
\includegraphics[width=0.98\linewidth]{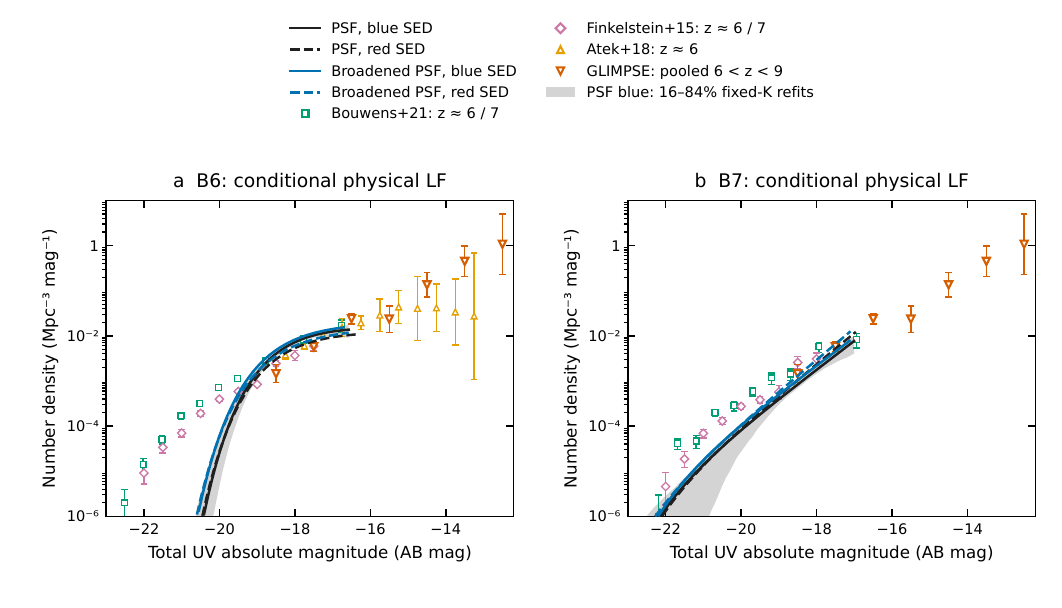}
\end{center}
\caption{\revised{Comparison of the four physical Schechter LFs with
published determinations. Our curves stop at each
calibration's true-luminosity support boundaries. Grey shading shows the
pointwise 16th--84th percentile fixed-response null-refit distribution for
PSF/blue SED and excludes uncertainty in calibration and population assumptions.
Open symbols show \citet{Bouwens2021} and \citet{Finkelstein2015} at
$z\simeq6$ and 7, \citet{Atek2018} at $z\simeq6$ and the pooled GLIMPSE
measurements \citep{Atek2026}, with their published values, uncertainties
and native cosmologies. Some error bars are smaller
than the symbols, and nonpositive lower endpoints are clipped by the
logarithmic axis. The same pooled $6<z<9$ GLIMPSE points
appear in both panels.}\\
{\revised{Alt text: Two luminosity-function panels showing four model
curves and open published comparison symbols on logarithmic density axes.
The present curves do not extend to the much fainter GLIMPSE points.}}}
\label{fig:literature}
\end{figure*}

\revised{Figure~\ref{fig:literature} compares fitted physical LFs with
published measurements. GLIMPSE probes Abell~S1063, combining
$6<z<9$, and finds a faint-end slope $\alpha=-1.98$ with no clear turnover
down to $M_{1500}\simeq-12.3$ \citep{Atek2026}, beyond our calibrated range.
Our four conditional nulls give B6-to-GLIMPSE native-window density ratios of 2.2--3.4
over $[-19,-18]$ and 1.3--2.0 over $[-18,-17]$. For B7, the ratio over
$[-19,-18]$ is 0.63--1.00. The next whole window is not supported by all
four responses. These ratios give context only and are not a tension test,
because the two analyses differ in redshift coverage, sightline, selection,
galaxy sizes, lens models and native cosmology.}

\revised{B6's adjacent-window mean densities imply a power-law index of
$-1.97$ to $-1.89$, close to GLIMPSE's fitted $\alpha=-1.98$.
GLIMPSE's two binned densities imply a steeper $-2.5\pm0.6$ (from their
quoted errors), so the B6-to-GLIMPSE density ratio decreases from the brighter window to
the fainter one. Neither average describes the local B6 shape.
For a Schechter LF per unit luminosity, the local slope is
$\alpha_{\rm loc}(M)=\alpha-10^{-0.4(M-M_*)}$. In the fiducial B6 fit it
changes from $-3.66$ at $M=-19$ to $-1.31$ at $M=-17$, whereas the GLIMPSE
fit stays near $-2$.}

\revised{We test the shape difference by fixing B6 to the GLIMPSE
$(M_*,\alpha)=(-21.29,-1.98)$, fitting its normalization and all three B7
parameters. Relative to the six-parameter free Schechter fit,
$\mathrm{AIC}_{\rm fixed}-\mathrm{AIC}_{\rm free}=+12.31$ to $+15.85$.
Removing the eight B6 proxy bins brighter than $-19$, containing six
sources, changes this to $-2.35$ to $-1.46$. The remaining counts weakly
distinguish a faint knee/shallow slope from a bright knee/steep slope.
The adopted GLIMPSE shape includes the blank-field bright end of
\citet{Bouwens2021}. All four B6 nulls agree with the \citet{Bouwens2021}
$z\simeq6$ points to within 0.2\,dex at $M\simeq-18.8$ and $-17.8$, but lie
1.1--1.5\,dex below them at $M\simeq-20$ (figure~\ref{fig:literature}a).
The B6 knee is therefore set by the scarcity of bright sources. At these
magnitudes the optical veto lowers the B6 response,
and the omitted source shot noise and deblending matter most (Supplementary
Data). We attribute the knee to this sample's bright end rather than to the
$z\simeq6$ LF, and the fitted B6 $\alpha$ describes the knee and the faint
slope together (Supplementary Data). The curvature test compares B6 only with
its own Schechter fit. The B6 and GLIMPSE shapes still differ.}

\section{Discussion}
\label{sec:discussion}

\revised{Relative to the earlier preprint, we have changed both the
observation model and the model used for comparison. The flux convention,
rest-frame magnitude mapping, selection response, source-plane volume and
bin-integrated count expectation have all changed, and $M_*$ is now free
rather than fixed at $-20.5$. The earlier magnitude calculation omitted
the AB bandwidth term $+2.5\log_{10}(1+z)$, amounting to $2.11$--$2.39$ mag
over $z=6$--8. Because several parts of the calculation have changed,
isolated comparisons would be needed to attribute the change in model
ranking to any one of them.}

The unrestricted Schechter null is essential. \revised{With $M_*$ free, its
knee can supply faint-end flattening without any extra turnover factor.
The fiducial B6 null, for example, flattens from $\alpha_{\rm loc}=-3.66$ at $M=-19$ to
$-1.31$ at $M=-17$ (subsection~\ref{sec:comparison}), although its density
per magnitude still rises up to the support boundary.}
A positive $\Delta\mathrm{IC}$ therefore says only that the calibrated data do
not justify two additional curvature parameters.  \revised{It does not
exclude a turnover of any shape, because the tested cases of
equation~(\ref{eq:curvature}) are detected in only 5--26 per cent of
simulated trials (subsection~\ref{sec:sensitivities}). Nor does it
constrain luminosities fainter than the five-per-cent support boundaries of
subsection~\ref{sec:sampleresponse}.} \revised{HFF analyses of magnification and completeness systematics
\citep{Atek2018,Bouwens2022a,Bouwens2022b} and newer JWST cluster studies
\citep{Atek2026,Goolsby2026} reach fainter luminosities than our response
support. Our comparison with GLIMPSE (subsection~\ref{sec:comparison})
therefore covers only the overlapping range.}

The absolute normalization remains more uncertain than \revised{counting noise
alone implies}. Lens-model uncertainty is not contained in the fixed-best-map
bootstrap intervals. \rtwo{Within the UNCOVER model family, the rescaled
model-range maps give a statistical magnification uncertainty of 0.01\,mag
(median) and at most 0.21\,mag for our sources. Independent lens models
differ more. For the matched sources, GLAFIC magnitudes are 0.24\,mag
brighter than UNCOVER ones (median; Supplementary Data).}
\revised{Independent lens families and assumed contaminant populations can
change both normalization and shape.}
Further limits
come from a single cluster sightline, the two injected SEDs and \revised{prescribed
image profiles}, possible unregistered counterimages, residual low-redshift
interlopers, and recovery in crowded regions.  Cosmic variance is not included
in the tabulated counting intervals \citep{Trenti2008,Robertson2010,Trapp2020,
Dawoodbhoy2023}.  \revised{Our model comparison applies to this sample,
its observing conditions and its supported luminosity range. It does not
establish the cosmic-mean faint-end form. The spectroscopically established
$z=7.88$ overdensity behind the cluster \citep{Morishita2023} illustrates
this concern. We retain that region and do not classify its photometric
neighbours as confirmed members.}

\section{Summary}
\label{sec:summary}

\revised{Our seven-band injection response connects the $z=6$--8 UVLF
behind Abell~2744 to aperture counts and combines counterimages in the source
plane. Of 184 sources, 151 enter 25 bins common to four compact-profile/SED
calibrations. Their five-per-cent relative-peak boundaries are approximately $-16.4$
to $-16.7$ in B6 and $-17.0$ to $-17.1$ in B7. Adding curvature does not improve the B6 fit
($\Delta\mathrm{AIC}=+4.00$, the two-parameter penalty alone) and barely
improves B7 ($\Delta\mathrm{AIC}=+3.87$ to $+3.95$). The B6 null itself has a faint knee
($M_*\approx-17.9$ to $-18.2$, $\alpha\approx-0.9$ to $-1.0$;
Supplementary Data). Its
disagreement with the GLIMPSE shape comes from the bins brighter than
$-19$, which hold six sources.
The matched-area GLAFIC test leaves the curvature ranking unchanged.
The tested suppressions reduce density by 21 or 60 per cent one magnitude
beyond onsets at $M_{\rm t}=-18.5$ or $-17.5$.
With the fiducial PSF/blue-SED response held fixed, simulations detect
these cases in only 5--26 per cent of trials, using null-calibrated detection
thresholds with a 5 per cent false-positive rate.}
\revised{Within the stated response assumptions, these Abell~2744 data do
not require additional faint-end curvature. A steep assumed B6 contaminant
template near 30 per cent can weakly reverse the ranking, depending on the
curvature bound.}

\section*{Funding}
The author received no specific funding for this work.

\begin{ack}
The author thanks the anonymous referee for a careful report. The referee's
questions on how strong a turnover must be to be detected and on the
normalization of the public UNCOVER model-range lens maps improved this paper.
This work uses public products from the JWST UNCOVER Treasury survey (Program
2561; PIs Labb{\'e} and Bezanson) and the MegaScience survey (Program 4111; PI
Suess).  It is based on observations made with the NASA/ESA/CSA James Webb
Space Telescope and products distributed through the Mikulski Archive for
Space Telescopes at the Space Telescope Science Institute.  The author thanks
the survey and lens-model teams for releasing their data products, and
L.~Furtak and A.~Zitrin for confirming the normalization of the
model-range lens maps.
\revised{The independent comparison uses GLAFIC mass maps provided through
the Hubble Frontier Fields program and MAST.}
\revised{OpenAI Codex assisted with writing part of the analysis code.  The
author accepts full responsibility for the analysis and manuscript.}
\end{ack}

\section*{Data availability}
The public photometric catalog, spectroscopy, and lens products are available
through the UNCOVER releases
\citep{UNCOVERDR3Data,UNCOVERSpectroscopyData,UNCOVERLensData,UNCOVERDR4}.
\revised{The independent GLAFIC v4 maps are identified by their versioned
MAST HLSP product series \citep{GLAFICv4Data}.}
\revised{The accompanying data and code archive provides the selected-source
table, calibrated forward kernels, fit results, simulation draws,
sensitivity tables and numerical figure data. The supplied kernels and
portable scripts reproduce the primary joint fits and core result figures.
Selected upstream scripts document the image-level calibration, which is
outside the scope of this portable entry point. The full image-level pipeline
is available from the corresponding author on reasonable request.}

\section*{Software}
The analysis uses Python, NumPy \citep{Harris2020}, SciPy \citep{Virtanen2020},
Astropy \citep{Astropy2022}, Matplotlib \citep{Hunter2007}, and EAZY
\citep{Brammer2008}. \revised{Source detection and aperture measurements use
the SEP library \citep{Barbary2016}, which implements algorithms from
Source Extractor \citep{BertinArnouts1996}. We use SEP rather than running
the standalone Source Extractor executable.}

\bibliographystyle{pasj-astronomy}
\bibliography{references}

\clearpage
\onecolumn
\section*{Supplementary Data}
\setcounter{section}{0}
\setcounter{table}{0}
\setcounter{figure}{0}
\setcounter{equation}{0}
\renewcommand{\thesection}{S\arabic{section}}
\renewcommand{\thetable}{S\arabic{table}}
\renewcommand{\thefigure}{S\arabic{figure}}
\renewcommand{\theequation}{S\arabic{equation}}

This Supplementary Data document reports the calibration and supporting
analyses for the primary model comparison. \revised{The observed quantity
is the fixed-PSF-corrected aperture proxy $\widetilde M$. The forward
response maps physical total $M_{1500}$ into that proxy, including
redshift and flux migration.} It uses the same joint Poisson likelihood
and curved-minus-Schechter sign convention as the main paper. Positive
$\Delta\mathrm{AIC}$ or $\Delta\mathrm{BIC}$ favors the Schechter model.

\section{\revised{Catalog quality and photometric selection}}

\revised{Table~\ref{tab:qualityflags} lists the released catalog flags used
in the revision. BCG denotes a bright cluster galaxy.
We retain these flags from the release, because refitting the
seven-band data does not regenerate them. Of the new 261 selector-passing
candidates, 22 fail the old seven-gate rule. \texttt{use\_phot} rejects 19
of them, artifacts 19, near-BCG 13 and the old EAZY flag 9, and the other
three flags reject none. These failures overlap, so we neither sum the
counts nor use them to correct the area. Dropping only the old
EAZY requirement leaves 242 candidates. Three of these fail only that
requirement but have successful new fits. Two of their old flag values are
missing sentinels and one is zero. We use their new posterior eligibility in
place of the quicklook flag and retain the seven-gate sample as a
sensitivity check.}

\begin{table}[!htbp]
\begingroup

\tbl{\revised{Released D032 catalog flags and their role in the revision.}}{%
\begin{tabular}{lll}
\toprule
Column & Revision requirement & Meaning in the public release \\
\midrule
\texttt{use\_phot} & 1 & Reliable catalog photometry \\
\texttt{flag\_nophot} & 0 & No viable aperture photometry when set \\
\texttt{flag\_lowsnr} & 0 & Detection-stack S/N below 3 when set \\
\texttt{flag\_star} & 0 & Stellar classification when set \\
\texttt{flag\_artifact} & 0 & Bad pixel, image artifact or subtraction residual \\
\texttt{flag\_nearbcg} & 0 & Within 3 arcsec of a subtracted BCG when set \\
\texttt{flag\_eazy} & 1 in sensitivity only & Reliable original quicklook fit \\
\bottomrule
\end{tabular}}
\label{tab:qualityflags}
\endgroup
\end{table}

\revised{The preliminary screen requires valid F814W and the two bands of
at least one S/N route, signed F814W S/N $\leq2$, and the corresponding
5/3 near-infrared thresholds. We fit photometric redshifts only for objects
passing either route. The final selector combines these requirements with
$p_{68}=P(6\leq z<8)\geq0.6$ and $p_{68}/P(z<6)\geq3$.
It assigns B6 when $p_6\geq p_7$ and B7 otherwise, independently of which
S/N route admitted the object.}

\revised{The fitting library has 15 nonnegative SFHZ+AGN components
(\texttt{agn\_blue\_sfhz\_13.param}). The two injected stellar templates
are its first two components, not an independent library. This calibration
tests recovery for those SEDs but not mismatch to an independent SED family.
EAZY \texttt{0.6.5.dev1+g9809a6c} uses 607 nodes
$z_j=\exp[\ln(1.01)+0.005j]-1$, $j=0,\ldots,606$, spanning
$0.01\leq z\leq19.9042$. We add a five-per-cent flux floor to the empirical
error once: $\sigma_{\rm eff}^2=\sigma_{\rm emp}^2+(0.05|f|)^2$.
The template-error curve \texttt{template\_error\_10.def}, scaled by 0.20,
gives $V_b(z)=\sigma_{{\rm eff},b}^2+[\mathrm{TEF}_b(z)\max(f_b,0)]^2$.
The log likelihood includes both $-\chi^2/2$ and
$-\tfrac12\sum_b\ln V_b$ over valid bands. We use no magnitude or UV-slope
prior or additional zeropoint offsets. Weighted template columns are
normalized before nonnegative least squares. Bounded-variable least squares
is used only if the 2000-iteration limit is reached. We normalize $P(z)$
by trapezoidal integration in $dz$ and integrate its piecewise-linear
interpolant for interval probabilities. Invalid bands are excluded, at
least two valid bands are required, and valid negative fluxes are retained.
The archive gives the complete configuration and portable seven-band
examples for one real and one artificial source, including posterior and
conditional luminosity-proxy refits. Detection, flags and lens quantities
are inputs to these examples, not reconstructed survey products.}

\revised{We define the clean region by dilating catalogue segmentation
labels carrying any stellar, artifact or near-BCG flag with the integer disk
$\Delta x^2+\Delta y^2\leq16$ on the native 40-mas grid.  A rounded origin0
physical coordinate must have all queried disk pixels inside the image and
none of their labels in the bad-label list. This pixel rule does not compute
exact fractional aperture overlap or reconstruct the released star-spike/BCG masks.
Among the 261 selector candidates, it admits the same 242 rows as the six
retained flags. The all-four native-weight-supported domain contains
567,300 original rays and retains all current source representatives.
Of 256 sampled backgrounds, 251 lie inside, with at least five in each of
the 36 design strata. We use their original inverse inclusion probabilities
for the conditional calibration and retain failed visits in the denominators.
Recovered fake centroids must also pass this domain. We apply support after
constructing the original lens-branch topology and before counterimage union
and source-plane integration.}

\section{\revised{Targeted spectroscopic check}}

\revised{The revised photometry-only source table has 14 unique associations
with successful public DR4.1 spectra of quality 2 or 3, of which 12 have
quality 3 and two have quality 2.
We match image identifiers to \texttt{id\_DR3} and de-duplicate them with
the current source registry. The three objects admitted
by replacing the old quicklook-EAZY flag have no association in the
668-row public DR4.1 table.  Table~\ref{tab:specoutside} lists the four
catalogue best-fit redshifts outside $6\leq z<8$.
Removing them leaves 184 sources (124 B6, 60 B7) before fitted-luminosity
range selection. Removing only quality-$\geq3$ objects leaves 185
(124/61), and retaining all spectra gives 188 (127/61).}

\begin{table}[!htbp]
\begingroup

\tbl{\revised{Catalog-best spectroscopic redshifts outside the target
interval. The fixed-aperture proxies $\widetilde M$ use conditional
high-redshift photometric estimates, not refits at the spectroscopic redshifts.}}{%
\begin{tabular}{rrrrr}
\toprule
DR3 id & Best $z_{\rm spec}$ & Quality & Photometric bin & $\widetilde M$ \\
\midrule
35563 & 5.794448 & 3 & B6 & \revised{$-18.742$} \\
35885 & 5.990177 & 3 & B6 & \revised{$-16.607$} \\
48347 & 1.416880 & 2 & B7 & \revised{$-18.297$} \\
54761 & 5.666573 & 3 & B6 & \revised{$-17.667$} \\
\bottomrule
\end{tabular}}
\label{tab:specoutside}
\endgroup
\end{table}

\revised{The catalog best redshift and template-fit percentiles can come
from different methods, so the percentiles need not describe an
error interval around the best redshift. Object 35885 lies close to the
boundary and carries a local-background advisory. The full ledger retains
both methods, original spectral identifiers, matching distances and duplicate
choices. We report the effect of quality 2 separately because it denotes
solid/probable identifications rather than the secure quality-3 category.
With non-random targeting, neither four of 14 nor $\sum(1-p_{68})$ is an
independently calibrated whole-field contamination rate.}

\section{\revised{Seven-band artificial-source experiment}}

\revised{We render source light into all seven bands at every redshift,
rather than correcting F814W alone as in the earlier experiment.
Table~\ref{tab:injectiondesign} gives the design. The two public stellar
SFHZ templates include nebular emission and differ in nominal attenuation,
$A_V=0.01$ and 0.25 (\texttt{corr\_sfhz\_13\_bin0\_av0.01.fits}
and \texttt{corr\_sfhz\_13\_bin0\_av0.25.fits}, EAZY
\texttt{0.6.5.dev1+g9809a6c}). Their spectra are redshifted through each observed
filter with Inoue intergalactic absorption, and normalized to the specified
F150W magnitude.  The rest-1500 and observed-filter integrals use the same
template state. The five prescribed circular profiles test sensitivity
to image-plane convolution width. Their widths are neither inferred
intrinsic galaxy radii nor a population size distribution.}

\begin{table}[!htbp]
\begingroup
\tbl{\revised{Primary artificial-source design.  All conditions reuse the
same independent background positions.}}{%
\begin{tabular}{ll}
\toprule
Quantity & Values \\
\midrule
Background positions & 256; 251 within the adopted clean domain \\
Redshifts & 6.125, 6.375, 6.625, 6.875, 7.125, 7.375, 7.625, 7.875 \\
F150W magnitudes & 19, 21, 23, 25, 26, 27, 27.5, 28, 28.5, 29, 29.5, \\
 & 30, 30.5, 31, 33, 36, 39 \\
\rtwo{Stellar-population SEDs} & Nominal $A_V=0.01$ and 0.25 \\
Image profiles & PSF; PSF convolved with circular Gaussians \\
Gaussian widths & $\sigma=0.05,0.10,0.20,0.40$ arcsec in the image plane \\
Total visits & 348,160; 341,360 in the clean-domain calibration \\
\bottomrule
\end{tabular}}
\label{tab:injectiondesign}
\endgroup
\end{table}

\revised{In each visit we add one deterministic signal to the unmodified
local background, run native long-wave detection, and measure PSF-matched
colors at the recovered centroid. We then apply the same signed optical
veto, near-infrared S/N routes and seven-band posterior calculation. We also
apply the long-wave cut that reproduces the catalog's \texttt{flag\_lowsnr}
(table~\ref{tab:qualityflags}), a signed S/N $\geq3$ in the PSF-matched
F277W+F356W+F444W aperture. On the released photometry this cut matches
\texttt{flag\_lowsnr} for all 74,020 D032 entries. We do not add a new source-shot-noise realization
or rerun the full-mosaic catalogue's deblending and flag construction.
These approximations matter most for bright sources and crowded or
extended objects.}

\revised{Detection uses the native noise-equalized F277W/F356W/F444W
stack and SEP 1.2.1: threshold 1.2, minimum area three pixels, a
$13\times13$ Gaussian matched filter with $\sigma=3.5/2.35$ pixels,
32 deblending levels, contrast $10^{-4}$ and cleaning disabled. Recovery
requires a unique match within two pixels of the injected position.
Ambiguous matches and mergers with pre-existing detections are rejected.
The PSF-matched long-wave aperture cut is applied separately.}

\revised{Across the full design, 167,245 visits yield a new matched
detection and 57,838 pass the detection/color/long-wave selector.  Of the
latter, 57,492 return to their input redshift interval and 346 enter the
other interval.  Restricting truth positions to the adopted domain retains
57,127 accepted visits, and requiring the recovered centroid to lie in that
same domain rejects another 215.  The resulting 56,912 acceptances contain
56,588 same-interval and 324 cross-interval recoveries. Because of unequal sampling
probabilities and repeated backgrounds, these are design counts, and
their ratios do not estimate population completeness or contamination.}

\revised{We compared local remeasurements of 16 preselected catalog sources
with their released D032 aperture measurements at fixed catalog centroids.
The sample spans three magnitude strata and a quality-rejected stratum from
the initial selection, and 13 of its objects enter the revised 184-source sample.
After PSF matching and Galactic correction, all 112 seven-band flux residuals
are below $6.1\times10^{-4}$ times the released aperture error.
Positive-flux adjacent-band colors agree within $1.8\times10^{-4}$ mag, and
the reconstructed aperture errors agree to numerical precision.
The optical veto, both near-infrared S/N routes and long-wave S/N flag give
identical outcomes for all 16 sources. All pass the preliminary screen and
long-wave threshold, so this set does not test rejection near these boundaries.
The error coefficients come from the public catalog, and one object's weight
normalization was checked against the other 15.
These sources also informed the measurement conventions, so this comparison
checks the implementation. The next paragraph tests 30 held-out objects.
The archive provides the
individual measurements and comparison code.}

\revised{We froze 30 further catalogue objects before examining their
remeasurements, excluding the 16-source development set and covering
threshold boundaries, quality rejections and crowded/isolated regions in
27 cached calibration cutouts. Local redetection recovered all 30, and
their S/N cuts and posterior-based membership were unchanged. One artifact-flagged
object shifted by 0.44 pixels, giving a maximum flux difference of
38.1 times its catalogue error while remaining rejected. For the seven
accepted objects, full posterior and conditional-eight-node SED refits,
with magnification reevaluated at the conditional mean redshift, preserved
all proxy bins (six in the fitted domain). The largest proxy change was
$6.7\times10^{-6}$ mag. Quality flags and geometric lens representatives
were inherited.}

\section{\revised{Spatial response and flux recovery}}

\revised{For each profile, SED, redshift and apparent magnitude, we estimate
local recovery from a weighted average of binary outcomes at independent
backgrounds in the same foreground-environment class. Gaussian weights in
standardized observing-depth coordinates multiply the original inverse
inclusion probabilities and are normalized locally. Whole-background
cross-validation selects F814W and F150W depth predictors with 12 nearest
neighbours from comparisons with a four-band predictor and 12--64 neighbours.
We exclude all conditions of each held-out background together and interpolate
linearly within the measured magnitude grid. The response can be non-monotone
because the optical non-detection cut can reject bright objects.}

\revised{We fit eight conditional redshift nodes for all 57,838 selected
visits. The initial artificial photometry uses a known-profile correction
$q_p$ common to all bands. For the primary observable we replace it by the
same fixed PSF correction $q_0=1.550166015521632$ used for real sources:
$\widetilde M=M_{\rm known\,profile}+2.5\log_{10}(q_p/q_0)$.
The real-source transformation instead uses the released $q_{\rm cat}$.
We multiply all fluxes and errors of an object by the same positive scalar
and retain its valid-band mask. With the homogeneous error model and no-prior
likelihood, this preserves colors, S/N, $P(z)$ and membership. Fixed-redshift
refits of six sources at three redshifts confirm the scaling, with a maximum
$\chi^2$ change below $1.3\times10^{-14}$. This tests the change of flux
units rather than SED accuracy.}

\revised{The primary response uses every selected visit without requiring
a measured Kron correction. Its mapping from true total luminosity to the
proxy therefore retains profile-dependent aperture loss. In a separate diagnostic,
48,532 selected visits have a local public-Kron measurement and 9306 do not.
Enlarging cutouts for the 1647 initially unresolved compact-profile visits
makes 1380 computable, while 267 still require nonlocal neighbour or identity
information. Because larger cutouts can change the deblending context,
these measurements test sensitivity rather than replay the full catalog
exactly. We do not substitute perfect-flux outcomes for unresolved measurements
or changes of context.}

\subsection{\revised{Forward fit and supported observations}}

\revised{The four compact responses use the PSF and the PSF convolved with
a $\sigma=0.05$-arcsec Gaussian, each with both SEDs. They are evaluated at
101 true-total-magnitude knots over $-23\leq M\leq-13$, eight true-redshift
nodes, and 46 recovered categories:
20 half-magnitude bins in each assigned interval, plus brighter, fainter
and undefined-proxy categories. For each branch we mix the joint empirical
selection, assigned interval, F150W S/N and magnitude residual from the
12 local backgrounds and adjacent apparent-magnitude nodes, retaining
failures in the probability law. Outcomes must pass the clean domain and
$\mu\leq30$ at their recovered conditional mean redshift before the
highest-F150W-S/N branch is chosen. We retain migration in both directions
across the magnification cut. Under conditional independence of branches,
the sum over winner categories recovers the counterimage union. Source-cell
area and the redshift shell enter once, without an additional $1/\mu$
factor. We evaluate lens quantities at the geometric representative and
do not propagate the lens gradient across the recovered centroid displacement.}

\revised{For the fiducial PSF/blue-SED null, cells with multiple active
branches contribute 1.69 and 2.43 per cent of the expected fitted B6 and B7
counts. In eight public-image patches at two redshift nodes, refining the
image/source sampling changes the summed retained-bin response by at most
4.3 per cent over $M=-20,-18,-17$. The patches are not a representative
area sample, and individual weak tails can differ substantially. Coarsening
the completed response's true-$M$ knots from 0.1 to 0.2 mag changes retained
expected counts by at most 0.5 per cent for the null and eight power-test
models, and by at most 3.4 per cent in bins with expectation at least one.
}

\revised{We separately added true-redshift nodes $z=6.25,6.50,6.75$ for
the fiducial PSF/blue-SED response, using 12,801 additional seven-band
visits at the same 251 backgrounds and 17 apparent magnitudes. We repeated
detection, posterior fitting and conditional luminosity-proxy measurement and
rebuilt the full-field source-plane response at each new redshift. We then
combined the seven B6 nodes with comoving shells bounded by adjacent-node
midpoints and the endpoints 6 and 7. B7 sampling was unchanged. At the original null parameters,
retained-bin predictions change by at most 0.96 count or
$0.18\sqrt{\lambda_k}$, and by 3.7 per cent where $\lambda_k\geq1$.
The common 25-bin mask and 151 observed counts are unchanged, and the faint
B6 true-magnitude support shifts by 0.0042 mag. Refit
$\Delta\mathrm{AIC}$ remains 4.00 for B6 and changes from 3.93 to 3.90
for B7. The weakly identified B7 curved-model parameters shift substantially
despite this small likelihood change. This check supports the model ranking
under the refined B6 redshift sampling.}

\revised{We measure observation-bin exposure by integrating the response
against a unit luminosity function across both true-redshift intervals.
The intersection of bins reaching five per cent of their recovered
interval's maximum in all four compact scenarios is
$-23\leq\widetilde M<-16.5$ in B6 and
$-23\leq\widetilde M<-17$ in B7. These 25 half-magnitude bins contain
151 of the 184 representatives, 114 in B6 and 37 in B7. The range is chosen
without using observed counts and includes zero-count bins. The threshold
measures relative exposure, rather than five-per-cent completeness for an
individual galaxy. All contributing responses lie within the measured
apparent-magnitude grid and have defined primary proxies, without response
extrapolation. We integrate the piecewise-linear response over the full
true-$M$ domain using 12-point Gauss--Legendre quadrature per knot interval,
rather than limiting the integral to the observed fitted bins. The
true-$M$ support and proxy-bin range can therefore have different boundaries.}

\revised{Table~\ref{tab:peakboundaries} gives the faint boundaries where
the response into retained bins, $R_t(M)=\sum_{k\,\mathrm{retained}}K_{kt}(M)$,
crosses 5 or 50 per cent of its own maximum. Crossings are interpolated
linearly between true-magnitude knots. In the fitted nulls,
the brightest one-magnitude interval of the full integration domain
contributes less than $9\times10^{-5}$ of retained counts, and the faintest
one-magnitude interval contributes zero.}

\begin{table}[!htbp]
\begingroup
\tbl{\revised{Faint true-total-magnitude boundaries of the response into
the primary 25 observed bins.}}{%
\begin{tabular}{llrrrr}
\toprule
Profile & $A_V$ & B6, 5\% & B6, 50\% & B7, 5\% & B7, 50\% \\
\midrule
PSF & 0.01 & $-16.57$ & $-17.59$ & $-17.01$ & $-18.14$ \\
PSF & 0.25 & $-16.39$ & $-17.41$ & $-16.96$ & $-18.38$ \\
Broadened PSF & 0.01 & $-16.71$ & $-17.71$ & $-17.13$ & $-18.28$ \\
Broadened PSF & 0.25 & $-16.57$ & $-17.53$ & $-17.10$ & $-18.53$ \\
\bottomrule
\end{tabular}}
\label{tab:peakboundaries}
\endgroup
\end{table}

\revised{Table~\ref{tab:compactnull} gives the joint Schechter point
estimates for this sample and observation mask. The four cases apply
different profile/SED assumptions to the same data. Each B6 curved
alternative returns the Schechter boundary with $\Delta\mathrm{AIC}=4$,
and the four B7 values are 3.87--3.95. We retain the final true-magnitude
interval $[-14,-13]$ in the forward integral even though its response to
the retained observations is zero in these cases.}

\begin{table}[!htbp]
\begingroup\small
\tbl{\revised{Joint Schechter estimates. The first four rows fit the
same 114 B6 and 37 B7 sources under the compact calibrations, with
$\phi_*$ in $\mathrm{Mpc}^{-3}$. The last three rows give marginal
quantiles of the 256 fiducial fixed-response null refits, not total
uncertainties or Bayesian credible intervals.}}{%
\begin{tabular}{llrrrrrr}
\toprule
Profile & $A_V$ & \multicolumn{3}{c}{B6} & \multicolumn{3}{c}{B7} \\
\cmidrule(lr){3-5}\cmidrule(lr){6-8}
 & & $\log_{10}\phi_*$ & $M_*$ & $\alpha$
   & $\log_{10}\phi_*$ & $M_*$ & $\alpha$ \\
\midrule
PSF & 0.01 & $-1.628$ & $-17.887$ & $-0.869$ & $-5.062$ & $-21.728$ & $-2.582$ \\
PSF & 0.25 & $-1.794$ & $-17.983$ & $-0.939$ & $-5.157$ & $-21.774$ & $-2.706$ \\
Broadened PSF & 0.01 & $-1.641$ & $-18.075$ & $-0.978$ & $-4.843$ & $-21.604$ & $-2.549$ \\
Broadened PSF & 0.25 & $-1.810$ & $-18.160$ & $-1.026$ & $-5.351$ & $-22.117$ & $-2.740$ \\
\midrule
\multicolumn{2}{l}{16th percentile} & $-1.837$ & $-18.259$ & $-1.313$ & $-6.748$ & $-24.000$ & $-2.772$ \\
\multicolumn{2}{l}{Median} & $-1.634$ & $-17.875$ & $-0.838$ & $-3.966$ & $-20.545$ & $-2.389$ \\
\multicolumn{2}{l}{84th percentile} & $-1.520$ & $-17.456$ & $-0.247$ & $-2.858$ & $-19.059$ & $-1.730$ \\
\bottomrule
\end{tabular}}
\label{tab:compactnull}
\endgroup
\end{table}

\revised{Both redshift intervals are fitted simultaneously. The common
numerical bounds are $-8\leq\log_{10}\phi_*\leq-0.3$,
$-24\leq M_*\leq-17$, and $-4.5\leq\alpha\leq2$.
Each single-interval curved alternative adds
$-21.5\leq M_{\rm t}\leq-13$ and $0\leq\beta\leq4$, while refitting
all six baseline parameters. The curvature factor begins at $M_{\rm t}$,
which need not be the maximum of the full luminosity function. We report
solutions on the numerical boundaries. At $\beta=0$, $M_{\rm t}$ is
unidentifiable. Count resampling and whole-background calibration resampling
measure different conditional uncertainties, both excluding lens-family
uncertainty and cosmic variance.}

\revised{The displayed central 68.27-per-cent Poisson intervals have a
zero-count upper endpoint of 1.841, equivalent to a one-sided
84.13-per-cent limit.}

\revised{All 256 null refits converge. In B7, $M_*$ reaches its lower
bound in 60 refits and its upper bound in three, and $\log_{10}\phi_*$
reaches its lower bound in five. These counts overlap. The lower $M_*$ percentile
is therefore truncated by the optimization domain, not a physical limit.
The archive includes parameter correlations and individual refits.}

\revised{We draw 32 whole-background resamples within the 36 design strata,
keeping repeated conditions together, the adopted neighbourhood model
fixed, and the same 151 observed counts in the 25-bin mask. For each
resample we rebuild the full eight-node, 101-magnitude response and refit
the six-parameter null and both eight-parameter alternatives. All 96 fits
converge. The $\Delta\mathrm{AIC}$ ranges are 3.994--4.000 in B6 and
3.726--3.996 in B7. The largest fractional sample standard deviations of
individual retained-bin exposures $E_k$ occur in the faintest bins:
12.0 per cent in B6 and 9.1 per cent in B7. At true $M=-18$, the
calibration-only sample standard deviations of refitted null
$\log_{10}\phi$ are 0.027 and 0.029 dex. These 32-resample summaries
describe calibration scatter only, separate from the Poisson refit
distributions.}

\revised{Expanding the shared $M_*$ bounds to $[-26,-16]$ for both models
leaves B6 at the null boundary and gives B7 $\Delta\mathrm{AIC}=3.86$--3.95.
Some curved fits reach the normalization bound instead, showing the
parameter degeneracy. A three-per-cent observation-exposure threshold
retains 162 sources (114/48), with no likelihood gain from curvature in
any compact scenario. A ten-per-cent threshold selects the same bins as five
per cent.
The two quicklook-flag policies combined with the three spectroscopic
exclusions give 155, 152 or 151 fitted sources. All favor the null under
the fixed PSF/blue-SED response, with B7 $\Delta\mathrm{AIC}=3.83$--3.93.
These comparisons change observed membership only, without recalibrating
the response for alternative populations.}

\subsection{\revised{Extended-profile full response}}

\revised{We reconstruct the full kernel for the existing
$\sigma=0.10$-arcsec Gaussian-convolved PSF trials, using both injected SEDs.
All 68,272 trial states at the 251 retained backgrounds enter, including
failures. The calculation uses the same eight redshift nodes, 101 true
magnitude knots, recovered-magnification cut and representative-image
rule as the primary response. Selection, F150W S/N, redshift assignment and
proxy-luminosity error are retained jointly. No new image injection is
performed.}

\revised{On the unchanged 25-bin mask and 151 counts, the six joint fits
give B6 $\Delta\mathrm{AIC}=4.00$ for both SEDs and B7 3.929 and 3.870
for $A_V=0.01$ and 0.25. The corresponding B6 null $(M_*,\alpha)$ values
are $(-18.358,-0.878)$ and $(-18.461,-0.956)$, and the B7 values are
$(-22.427,-2.643)$ and $(-22.284,-2.710)$. The archive includes the complete
kernels and a kernel-to-fit reproduction script.}

\rtwo{A population in which each profile/SED case has the same share has a
response equal to the mean of the individual kernels. On the same mask and
counts, the equal-weight mixture of the four primary calibrations gives B6
$\Delta\mathrm{AIC}=4.00$ and B7 3.915, with null $(M_*,\alpha)$ of
$(-18.02,-0.95)$ in B6 and $(-21.81,-2.65)$ in B7. Adding the two
$\sigma=0.10$-arcsec kernels gives B6 4.00 and B7 3.916, with
$(-18.07,-0.89)$ and $(-21.94,-2.66)$.}

\revised{All originally fitted faint bins pass the extended response's
five-per-cent exposure rule. Only the brightest B6 bin, $[-23,-22.5)$,
falls below it, with relative exposures 0.021 and 0.026 and zero observed
sources. A six-scenario common mask would therefore retain 24 bins but
the same 151 sources. The reported sensitivity keeps all 25 primary bins
so that it compares identical observations.}

\subsection{\revised{Bright-bin sensitivity of the B6 shape}}

\revised{We test which observations distinguish the B6 fit from the
GLIMPSE shape by excluding all eight B6 proxy bins with
$\widetilde M<-19$, including empty bins. The retained count vector has
108 B6 and 37 B7 sources in 17 bins. The full true-luminosity integral and
B7 observations remain unchanged. For each compact response we compare
the six-parameter joint Schechter fit with one fixing the B6 shape to
$M_*=-21.29$, $\alpha=-1.98$, retaining a free B6 normalization and all
three B7 parameters. This four-parameter comparison tests the shape within
our response. It does not fit GLIMPSE's counts or propagate its parameter
uncertainties. The adopted shape is the published GLIMPSE fit that also
includes the Bouwens et al. (2021) bright-end measurements. For the LF
per magnitude, $\phi(M)$, the local luminosity power-law index is
$\alpha_{\rm loc}=-1-(1/0.4)\,d\log_{10}\phi/dM$, not
$d\log_{10}\phi/dM$ itself.}

\revised{For the primary mask, fixed-GLIMPSE-shape minus free-Schechter
AIC is $+12.31$ to $+15.85$. After excluding the bright bins it is
$-2.35$ to $-1.46$, so a free B6 shape no longer earns its extra parameters.
The free-knee solutions then have $\alpha=-1.31$ to $-0.93$.
Fixing only $M_*=-21.29$ instead gives $\alpha=-1.90$ to $-1.76$ and loses
only $2\Delta\ln\mathcal L=0.60$--1.76 relative to the free knee.
Thus bright-bin information, not a resolved power-law faint tail, drives
the full-sample shape discrimination. All 24 fits and their
parameter counts are archived.}

\Needspace{5\baselineskip}
\subsection{\revised{Conditional detection power}}

\revised{Using the fiducial PSF, $A_V=0.01$ response and the four-scenario
common observation mask, we calibrate the statistic with 256 Poisson
realizations of the fitted joint Schechter model:
$T=2(\ell_{\rm curved}-\ell_{\rm Schechter})$. Detection requires $T$
strictly above the empirical 95th percentile of the corresponding
single-interval null distribution, which is 3.838 in B6 and 1.476 in B7. These
are separate tests, not a family-wise five-per-cent criterion or the
$\Delta\mathrm{AIC}<0$ rule. Each null distribution has 12/256 draws
above its threshold. Under the $\Delta\mathrm{AIC}<0$ rule the null rates are 11/256 (B6) and 0/256 (B7), and the eight cases give 12, 25, 5 and 9 (B6) and 0, 6, 1 and 3 (B7) detections in 128.}

\revised{These power calculations use only the fixed fiducial PSF,
$A_V=0.01$ response. Summing its full kernel over all retained observed
bins, including redshift migration, gives zero response for true B6
$M\geq-15.9$ and true B7 $M\geq-16.2$. With the adopted linear
interpolation, onsets at or fainter than those boundaries leave every
retained-bin prediction unchanged for any $\beta$. These boundaries mark the
edge of the finite kernel.
The tested onsets $-18.5$ and $-17.5$ are brighter and have the nonzero
powers below.}

\revised{Table~\ref{tab:curvaturepower} gives 128 realizations for each of
eight specified curved luminosity functions. We add curvature only to the
stated interval and keep the remaining parameters at their fitted null values,
without renormalizing the bright-end normalization. We draw Poisson counts
for all recovered categories and refit both models on the same observation
mask. All refits converge. The Wilson intervals give the Monte Carlo counting
uncertainty at the estimated thresholds and omit threshold and calibration
uncertainty. The detection probabilities of 4.7--25.8 per cent measure
discovery power for these generating models under one fixed response.}

\begin{table}[!htbp]
\begingroup
\tbl{\revised{Conditional power for the fixed fiducial PSF, $A_V=0.01$
response above the separate empirical null thresholds. Each row has 128 Poisson realizations, and the intervals are
95-per-cent Wilson Monte Carlo intervals. $M_{\rm t}$ refers to physical
total $M_{1500}$, not the aperture proxy.}}{%
\begin{tabular}{lrrrrr}
\toprule
Interval & $M_{\rm t}$ & $\beta$ & Detections & Power (\%) & Wilson interval (\%) \\
\midrule
B6 & $-18.5$ & 0.25 & 12/128 & 9.4 & 5.4--15.7 \\
B6 & $-18.5$ & 1.00 & 25/128 & 19.5 & 13.6--27.2 \\
B6 & $-17.5$ & 0.25 & 6/128 & 4.7 & 2.2--9.8 \\
B6 & $-17.5$ & 1.00 & 9/128 & 7.0 & 3.7--12.8 \\
B7 & $-18.5$ & 0.25 & 18/128 & 14.1 & 9.1--21.1 \\
B7 & $-18.5$ & 1.00 & 33/128 & 25.8 & 19.0--34.0 \\
B7 & $-17.5$ & 0.25 & 18/128 & 14.1 & 9.1--21.1 \\
B7 & $-17.5$ & 1.00 & 13/128 & 10.2 & 6.0--16.6 \\
\bottomrule
\end{tabular}}
\label{tab:curvaturepower}
\endgroup
\end{table}

\subsection{\revised{Assumed contaminant-count sensitivity}}

\revised{The calibrated response includes true redshifts only in
$6\leq z<8$. We test an additive contribution from outside this interval.
For each compact calibration, let $\lambda_k^0$ be its fitted baseline
prediction and $L_b^0=\sum_{k\in b}\lambda_k^0$ its total in retained
observed interval $b$. We replace the count expectation by $\lambda_k+c_k$, with}
\begin{equation}
\revised{c_k=fL_b^0\,
\frac{\lambda_k^0w_k}{\sum_{j\in b}\lambda_j^0w_j},\qquad k\in b.}
\end{equation}
\revised{We hold $c_k$ fixed while refitting both competing models, using
$f=0.05$, 0.10 or 0.20. These fractions refer to baseline expected counts,
not a fitted sample purity. The normalization control uses $w_k=1$ in both
observed intervals. A faint-weighted template uses
$w_k=10^{0.4(\widetilde M_k+18)}$, where $\widetilde M_k$ is the proxy-bin
center. The arbitrary pivot cancels in the normalization.
We apply this template either to both intervals or to B6 alone, setting
$c_k=0$ in B7 for the latter. The B6-only case probes a possible asymmetric
effect of migration across the lower redshift boundary, without modeling
that migration physically.}

\revised{All 36 nonzero combinations and four zero-contribution controls
use the same 151 integer counts, 25-bin mask, response kernels and parameter
bounds. All 120 null/alternative fits converge, giving
$3.71\leq\Delta\mathrm{AIC}\leq4.00$ across the tested cases.
The inferred LFs nevertheless change. At true $M=-18$, their null-fit
$\log_{10}\phi$ shifts by as much as about 0.10 dex, and the largest
absolute slope change is $|\Delta\alpha|=0.53$. The archive supplies every fit, count component and
parameter change. The curvature ranking is unchanged for these templates. They are
assumed inputs: the data do not measure the abundance or luminosity
distribution of objects outside $6\leq z<8$.}

\revised{We extend the B6-only faint weighting to
$w_k=10^{\gamma(\widetilde M_k+18)}$. The original $\gamma=0.4$ raises
the contaminant-to-baseline ratio by 2.51 per magnitude. A one-percentage-point
scan over $0\leq f\leq0.30$, followed by $f=0.35,0.40,0.50,0.60$,
retains positive B6 $\Delta\mathrm{AIC}$ (minimum 1.59) in all four
calibrations. B7 also retains positive values. The null Poisson deviance
ranges from 8.66 to 8.97 at $f=0.30$ and from 10.77 to 11.84 at $f=0.60$.
These values are descriptive and do not give a calibrated goodness-of-fit
probability.
Additional checks use $\gamma=0.2,0.8$ at $f=0.10,0.20,0.30$ and extend
the proportional and both-interval templates to $f=0.30$.}

\revised{The steeper $\gamma=0.8$ template raises the relative contribution
by 6.31 per magnitude. At $f=0.30$, it assigns about 34.2 expected B6
contaminants, and its faintest-bin component is 1.32--1.34 times that bin's
baseline prediction. Table~\ref{tab:contaminationr6} shows the resulting
weak B6 preference changes. All these steep-template curved optima reach
the chosen upper bound on $\beta$, even when it is raised from 4 to 8 or 16.
The fits retain
the same 151 counts, response, six baseline bounds and fixed $c_k$ in both
models.}

\revised{For the steep template with the original $\beta\leq4$ bound,
refining sampled sign changes places the first located
$\Delta\mathrm{AIC}=0$ crossings near $f=29.6$ per cent for PSF/blue
and 30.6 per cent for broadened-PSF/blue. Neither red-SED calibration
has a crossing at the sampled fractions up to $f=0.60$. These crossing
fractions depend on the template and the $\beta$ bound. The archive includes the scan, crossing brackets and bound checks.}

\begin{table}[!htbp]
\begingroup
\tbl{\revised{B6 curved-minus-Schechter $\Delta\mathrm{AIC}$ at an assumed
$f=0.30$ B6 contaminant contribution. The steep case is sensitive to the
numerical curvature bound. None of the fractions is an estimated
contamination rate.}}{%
\begin{tabular}{llrrr}
\toprule
Profile & $A_V$ & $\gamma=0.4$, $\beta_{\max}=4$
 & $\gamma=0.8$, $\beta_{\max}=4$ & $\gamma=0.8$, $\beta_{\max}=16$ \\
\midrule
PSF & 0.01 & $+3.42$ & $-0.05$ & $-0.58$ \\
PSF & 0.25 & $+3.93$ & $+1.16$ & $+0.82$ \\
Broadened PSF & 0.01 & $+3.57$ & $+0.08$ & $-0.41$ \\
Broadened PSF & 0.25 & $+3.93$ & $+1.07$ & $+0.72$ \\
\bottomrule
\end{tabular}}
\label{tab:contaminationr6}
\endgroup
\end{table}

\section{\revised{Conditional effective-volume support}}

\revised{Table~\ref{tab:conditionaldepth} gives an auxiliary diagnostic
before magnitude-migration treatment: the faint edges of half-magnitude
bins passing the five-per-cent effective-volume criterion in the full
UNCOVER domain. The criterion uses each curve's finite-grid maximum and
requires both bin edges and all interior volume knots to pass. The common
range across all five image profiles and both SEDs extends only from $-23$
to $-19.5$, too shallow for a useful shape-independent faint-end constraint.
The primary fit mask is defined from the forward response instead, and
the deeper PSF and broadened-PSF ranges apply only to those profiles.}

\begin{table}[!htbp]
\begingroup
\tbl{\revised{Faint bin edges from the conditional selection volumes,
before magnitude-migration treatment. Widths are image-plane Gaussian
convolutions of the PSF. All full-domain retained ranges begin at $-23$.}}{%
\begin{tabular}{lrrrr}
\toprule
Profile & B6, $A_V=0.01$ & B6, $A_V=0.25$ & B7, $A_V=0.01$ & B7, $A_V=0.25$ \\
\midrule
PSF & $-16.5$ & $-16.5$ & $-17.0$ & $-17.0$ \\
$\sigma=0.05$ arcsec & $-17.0$ & $-16.5$ & $-17.0$ & $-17.0$ \\
$\sigma=0.10$ arcsec & $-17.0$ & $-17.0$ & $-17.5$ & $-17.5$ \\
$\sigma=0.20$ arcsec & $-18.0$ & $-18.0$ & $-18.5$ & $-18.5$ \\
$\sigma=0.40$ arcsec & $-19.5$ & $-19.0$ & $-19.5$ & $-19.5$ \\
\bottomrule
\end{tabular}}
\label{tab:conditionaldepth}
\endgroup
\end{table}

\revised{Joint optical/near-infrared selection need not approach unity at
the bright end, and some bright bins in the smaller common UNCOVER/GLAFIC
domain also fail the criterion. We therefore retain explicit bin lists
rather than every bin brighter than a faint limit. These volumes integrate
source-cell detection unions, so the support threshold alone does not bound
flux-migration errors.}

\section{\revised{Candidate-based source-identity assessment}}

\revised{For all 184 current representatives, we evaluate all-pair
source-plane separations, nearest neighbours and finite covered-ray
counterimage regions. We follow up 18 pairs separated by at most 0.6
arcsec at either their own conditional mean redshifts or their arithmetic
common redshift, comparing posterior overlap, seven-band colors and image
parity. We inspect native images and segmentation for ambiguous configurations
and obtain continuous inverse-map solutions for selected cases. This finite
search, even with a small inverse-equation residual, neither measures lens
positional uncertainty nor guarantees an exhaustive image census.}

\revised{The local entries 19308/19309/19310 share image parity and lie in
aligned faint structure. The touching labels of 19309/19310 could be knots
or products of deblending rather than separate galaxies. The distant
opposite-parity entry 24575 has redshift/colors compatible with 19310, but
their magnification-corrected flux ratio and predicted image position depend
strongly on the choice of source centre. Care is needed to distinguish its
faint segment from the bright neighbour. We keep the public associations and
list duplicate-counting alternatives, without summing fluxes, in
Table~\ref{tab:identitycounts}.}

\begin{table}[!htbp]
\begingroup
\tbl{\revised{Source-identity hypotheses before the luminosity-range cut.
Alternative rows retain the highest-F150W-S/N representative 19309.
They do not provide joint or integrated galaxy photometry.}}{%
\begin{tabular}{lrrr}
\toprule
Counting hypothesis & Total & B6 & B7 \\
\midrule
Public associations only & 184 & 124 & 60 \\
Local three entries counted once & 182 & 124 & 58 \\
Also associate 24575 & 181 & 124 & 57 \\
\bottomrule
\end{tabular}}
\label{tab:identitycounts}
\endgroup
\end{table}

\revised{The two local removals affect the fixed-aperture proxy bin centred
on $-16.25$, and removing 24575 also affects the bin centred on $-16.75$.
Both lie outside the primary B7 fit range, leaving the 151 retained
observations unchanged. We track the ten published counterparts of systems
53, 84 and 86 individually. Four selected images represent three known
sources, one further image passes the new photometric selector but fails
the released quality cuts, and the other five fail the preliminary
photometric screen. An unselected predicted counterpart is therefore not, by itself,
evidence for a missing physical galaxy.}

\revised{Table~\ref{tab:identitypairs} summarizes all 18 close pairs,
including members that enter the primary fit. The unchanged 151 counts
above apply only to the two specified four-object counting alternatives.
Other local pairs retain unresolved galaxy-versus-knot identities.}

\begin{table}[!htbp]
\begingroup
\tbl{\revised{Candidate-pair checks for the 184-source sample. All entries retain their catalog identities.}}{%
\small\setlength{\tabcolsep}{3pt}
\begin{tabular}{lrrrrcll}
\toprule
DR3 IDs (A/B) & $d_\theta$ & $d_\beta$ & $O_z$ & $Q_{\rm col}$ & Parity & Fit A/B & Recorded evidence \\
 & (arcsec) & (arcsec) & & & A/B & & \\
\midrule
14109/14110 & 0.349 & 0.241 & 0.337 & 4.01 & $+/+$ & Y/Y & Local: unresolved \\
17792/17793 & 0.377 & 0.228 & 0.588 & 5.99 & $+/+$ & Y/Y & Local: unresolved \\
18029/18032 & 0.436 & 0.198 & 0.840 & 6.87 & $+/+$ & Y/Y & Local: unresolved \\
19110/19112 & 0.601 & 0.164 & 0.744 & 8.03 & $+/+$ & Y/Y & Local: unresolved \\
19308/19309 & 0.631 & 0.103 & 0.627 & 3.74 & $-/-$ & N/N & Local: unresolved \\
19308/19310 & 1.118 & 0.191 & 0.645 & 7.41 & $-/-$ & N/N & Local: unresolved \\
19308/24575 & 42.649 & 0.360 & 0.657 & 6.05 & $-/+$ & N/N & Distant: unresolved \\
19309/19310 & 0.487 & 0.088 & 0.632 & 4.82 & $-/-$ & N/N & Local: unresolved \\
19309/24575 & 42.187 & 0.361 & 0.651 & 2.70 & $-/+$ & N/N & Distant: unresolved \\
19310/24575 & 41.831 & 0.379 & 0.924 & 0.98 & $-/+$ & N/N & Distant: unresolved \\
19343/19485 & 1.347 & 0.419 & 0.619 & 2.62 & $+/+$ & N/Y & Distinct labels \\
20789/22696 & 12.999 & 0.579 & 0.093 & 7.45 & $-/+$ & N/N & Low $P(z)$ overlap \\
21417/21418 & 0.481 & 0.206 & 0.048 & 10.97 & $+/+$ & Y/Y & Low $P(z)$ overlap \\
22696/22697 & 0.357 & 0.173 & 0.692 & 7.72 & $+/+$ & N/N & Local: unresolved \\
25644/25718 & 0.491 & 0.192 & 0.630 & 8.82 & $+/+$ & N/N & Local: uncertain \\
37288/37289 & 0.259 & 0.076 & 0.350 & 6.70 & $+/+$ & Y/N & Local: unresolved \\
52532/52569 & 0.851 & 0.580 & 0.449 & 11.58 & $+/+$ & Y/Y & Local: unresolved \\
56919/56920 & 0.736 & 0.503 & 0.271 & 51.84 & $+/+$ & Y/Y & Colors differ \\
\bottomrule
\end{tabular}}
\label{tab:identitypairs}
\par\smallskip\noindent\parbox{0.98\textwidth}{\footnotesize
$d_\theta$ is the sky separation; $d_\beta$ and parity use the arithmetic mean of the two conditional redshifts in the fixed fiducial lens model.
Pairs enter the search when either the own-redshift or common-redshift source-plane separation is at most 0.6 arcsec.
$O_z=\int\min[p_A(z),p_B(z)]\,dz$ uses normalized full redshift posteriors.
$Q_{\rm col}=\Delta\mathbf{r}^{T}(C_A+C_B)^{-1}\Delta\mathbf{r}$ compares the six flux ratios to F150W; $C$ includes the linearized shared-denominator covariance and otherwise independent band errors.
These diagnostics are neither association probabilities nor calibrated significance tests.
Fit A/B indicates membership in the primary 25-bin likelihood using the current $\widetilde M$ values, not the catalog-total magnitudes.
Local entries can be knots, deblended structures or neighbouring galaxies. Distinct labels do not establish distinct physical sources.
Low $P(z)$ overlap ($O_z<0.1$) or a large color residual ($Q_{\rm col}>25$) disfavors a merge only as a screening diagnostic.
All associations remain unconfirmed; no new merge or flux summation is adopted.}
\endgroup
\end{table}

\revised{A further count-only test considers the six disjoint local pairs
14109/14110, 17792/17793, 18029/18032, 19110/19112, 52532/52569 and
37288/37289. For each pair, we retain the higher-F150W-S/N entry, breaking
an exact tie by the lower catalog ID. The retained entries are 14109,
17792, 18032, 19110, 52532 and 37288. This leaves 178 source representatives
and 146 fitted entries (110 B6 and 36 B7). The removed entry 37289 was
already outside the fitted range. With the same four response kernels, 25-bin mask
and parameter bounds, all 12 refits converge. The B6 alternatives give
$\Delta\mathrm{AIC}=+3.93$ to $+4.00$, and B7 gives $+3.70$ to $+3.82$.
These tests change only the count vector, without merging fluxes or
recalibrating the selection for integrated galaxies. The primary fit retains
the original 151 observations.}
\FloatBarrier

\section{\revised{Lens normalization and independent-model comparison}}

\rtwo{The public UNCOVER v2 README normalizes the convergence, shear and
deflection fields to $D_{\rm ds}/D_{\rm s}=1$. The best-fit maps follow this
convention: magnifications recomputed from the best-fit $\kappa$ and
$\gamma$ reproduce the released $z=8$ magnification map and the DR4
magnification catalogue to better than 0.2 per cent. Read with the same
convention, all 50 model-range maps are larger than the best fit by a nearly
uniform factor. Their enclosed convergence is 1.62--1.69 times the best-fit
value at every radius from 10 to 160 arcsec around the main BCG. At
$z_{\rm s}=9$ their tangential critical curves enclose an effective Einstein
radius of 82--87 arcsec, compared with 34.7 arcsec for the best fit. Traced
to the source plane, the 172 spectroscopic multiple-image rows scatter by
10.2--10.3 arcsec RMS in maps 001, 017, 033 and 050, compared with
0.25 arcsec for the best fit. In this test, distinct resolved features of
one system are centred separately.}

\rtwo{Multiplying $\kappa$, $\gamma$ and the deflections by 0.60 removes the
difference. The rescaled maps give Einstein radii of 34.5--35.3 arcsec, and a
single factor fitted to the multiple-image constraints is 0.597--0.601, with
source-plane scatter of 0.41--0.47 arcsec. The rescaled maps also reproduce
the DR4 catalogue's own posterior magnification percentiles object by object.
For 26,706 catalogue galaxies at $z>1$ with $1.2<\mu<15$, the 16th, 84th,
2.5th and 97.5th percentiles of $\mu$, $\mu_{\rm t}$ and $\mu_{\rm r}$ over the
50 rescaled maps match the catalogue's 68 and 95 per cent bounds with median
offsets below 0.001\,dex. The scale that best matches the catalogue's
$\mu$ bounds is 0.598--0.599. The release's uncertainties were therefore
computed with this normalization, and the raw model-range files exceed the
documented normalization by a factor of 1/0.60. This calibration uses only
the release's own products. The factor is the ratio of the 0.06 and
0.10-arcsec pixel scales. We confirmed with the UNCOVER lens modellers that
the pixel-scale conversion was omitted when the 0.1-arcsec model-range maps
were computed. The best-fit maps are unaffected, and corrected maps will
appear in a later release.}

\rtwo{With this rescaling, the model-range maps give our 184 sources a
statistical magnification uncertainty of 0.01\,mag (median 68 per cent
half-width), 0.026\,mag at the 84th percentile and at most 0.21\,mag. The
median over the maps differs from the best fit by $-0.015$ to $+0.015$\,mag
(16th--84th percentiles). This is small compared with the 0.5-mag proxy
bins, so the fiducial analysis keeps the best-fit reconstruction.
Differences between independent models are larger (below). The scripts and
per-map results accompany the data archive.}

\revised{We project the published UNCOVER and independent GLAFIC fixed
mass maps using the distance ratio in their original flat cosmology,
$H_0=70\,\mathrm{km\,s^{-1}\,Mpc^{-1}}$ and
$\Omega_{\rm m}=0.3$, with $z_{\rm l}=0.308$. Luminosity distances and
comoving-volume shells use Planck 2018. The mass reconstructions are not
refitted under Planck cosmology. Across the eight redshift nodes, switching
the lens distance ratio to Planck would change that ratio by at most
0.18 per cent.}

\revised{We compare the independent models within their common clean
image-plane domain of $7.63\,\mathrm{arcmin}^2$, rather than the full UNCOVER
field. Each model uses its own magnification and $\mu\leq30$ selection. Reselecting
from all 261 photometric candidates yields 80 UNCOVER representatives
(41 B6, 39 B7) and 79 GLAFIC representatives (41/38). Among the 79 shared
sources, the 16th, 50th and 84th percentiles of
$M_{\rm GLAFIC}-M_{\rm UNCOVER}$ are $-0.369$, $-0.236$ and $-0.049$ mag.
For the common-domain PSF, $A_V=0.01$ responses, we use each model's
source-cell geometry and convergence/shear with the same clean-domain
calibration and recovered-mean magnification convention. We retain the
original branch representatives after the common-domain cut. GLAFIC geometry
is unavailable outside its public finite-map footprint.}

\revised{The two common-domain responses define a shared
observation mask: $-22\leq\widetilde M<-16$ in B6 and
$-23\leq\widetilde M<-16.5$ in B7. This is distinct from the full-field
four-scenario mask. It contains 64 UNCOVER and 65 GLAFIC sources
(Table~\ref{tab:commonlens}). We determine membership independently for
each model, without forcing a joint magnification cap. Both
curvature alternatives return $\beta=0$ and $\Delta\mathrm{AIC}=4$
under either reconstruction.}

\begin{table}[!htbp]
\begingroup
\tbl{\revised{Matched-common-domain forward comparison for PSF,
$A_V=0.01$. $N_6$ and $N_7$ count sources in this comparison's common
observation mask, not the 114/37 full-field sample. Each
$\Delta\mathrm{AIC}$ compares an eight-parameter single-interval curved
alternative with the same lens model's six-parameter joint Schechter fit.}}{%
\begin{tabular}{lrrrr}
\toprule
Lens model & $N_6$ & $N_7$ & $\Delta\mathrm{AIC}_{\rm B6}$ & $\Delta\mathrm{AIC}_{\rm B7}$ \\
\midrule
UNCOVER & 37 & 27 & 4.00 & 4.00 \\
GLAFIC & 38 & 27 & 4.00 & 4.00 \\
\bottomrule
\end{tabular}}
\label{tab:commonlens}
\endgroup
\end{table}

\revised{Both common-domain B7 Schechter fits reach the bright numerical
boundary at $M_*=-24$. The B6 point estimates are $-17.12$ (UNCOVER)
and $-17.94$ (GLAFIC), with exponential bright tails weakly constrained
by zero-count bins. Extreme ratios of the fitted luminosity functions
in that tail cannot give a reliable uncertainty amplitude. We do not
rank absolute AIC values across the different model-specific count vectors.
The paired-domain calculation shows that the curvature ranking is
unchanged under two fixed public reconstructions.}

\end{document}